\input psfig

\catcode `\@=11 

\def\@version{1.3}
\def\@verdate{28.11.1992}


%
%
%
%
%
%

\font\fiverm=cmr5
\font\fivei=cmmi5	\skewchar\fivei='177
\font\fivesy=cmsy5	\skewchar\fivesy='60
\font\fivebf=cmbx5

\font\sevenrm=cmr7
\font\seveni=cmmi7	\skewchar\seveni='177
\font\sevensy=cmsy7	\skewchar\sevensy='60
\font\sevenbf=cmbx7

\font\eightrm=cmr8
\font\eightbf=cmbx8
\font\eightit=cmti8
\font\eighti=cmmi8			\skewchar\eighti='177
\font\eightmib=cmmib10 at 8pt	\skewchar\eightmib='177
\font\eightsy=cmsy8			\skewchar\eightsy='60
\font\eightsyb=cmbsy10 at 8pt	\skewchar\eightsyb='60
\font\eightsl=cmsl8
\font\eighttt=cmtt8			\hyphenchar\eighttt=-1
\font\eightcsc=cmcsc10 at 8pt
\font\eightsf=cmss8

\font\ninerm=cmr9
\font\ninebf=cmbx9
\font\nineit=cmti9
\font\ninei=cmmi9			\skewchar\ninei='177
\font\ninemib=cmmib10 at 9pt	\skewchar\ninemib='177
\font\ninesy=cmsy9			\skewchar\ninesy='60
\font\ninesyb=cmbsy10 at 9pt	\skewchar\ninesyb='60
\font\ninesl=cmsl9
\font\ninett=cmtt9			\hyphenchar\ninett=-1
\font\ninecsc=cmcsc10 at 9pt
\font\ninesf=cmss9

\font\tenrm=cmr10
\font\tenbf=cmbx10
\font\tenit=cmti10
\font\teni=cmmi10		\skewchar\teni='177
\font\tenmib=cmmib10	\skewchar\tenmib='177
\font\tensy=cmsy10		\skewchar\tensy='60
\font\tensyb=cmbsy10	\skewchar\tensyb='60
\font\tenex=cmex10
\font\tensl=cmsl10
\font\tentt=cmtt10		\hyphenchar\tentt=-1
\font\tencsc=cmcsc10
\font\tensf=cmss10

\font\elevenrm=cmr10 scaled \magstephalf
\font\elevenbf=cmbx10 scaled \magstephalf
\font\elevenit=cmti10 scaled \magstephalf
\font\eleveni=cmmi10 scaled \magstephalf	\skewchar\eleveni='177
\font\elevenmib=cmmib10 scaled \magstephalf	\skewchar\elevenmib='177
\font\elevensy=cmsy10 scaled \magstephalf	\skewchar\elevensy='60
\font\elevensyb=cmbsy10 scaled \magstephalf	\skewchar\elevensyb='60
\font\elevensl=cmsl10 scaled \magstephalf
\font\eleventt=cmtt10 scaled \magstephalf	\hyphenchar\eleventt=-1
\font\elevencsc=cmcsc10 scaled \magstephalf
\font\elevensf=cmss10 scaled \magstephalf

\font\fourteenrm=cmr10 scaled \magstep2
\font\fourteenbf=cmbx10 scaled \magstep2
\font\fourteenit=cmti10 scaled \magstep2
\font\fourteeni=cmmi10 scaled \magstep2		\skewchar\fourteeni='177
\font\fourteenmib=cmmib10 scaled \magstep2	\skewchar\fourteenmib='177
\font\fourteensy=cmsy10 scaled \magstep2	\skewchar\fourteensy='60
\font\fourteensyb=cmbsy10 scaled \magstep2	\skewchar\fourteensyb='60
\font\fourteensl=cmsl10 scaled \magstep2
\font\fourteentt=cmtt10 scaled \magstep2	\hyphenchar\fourteentt=-1
\font\fourteencsc=cmcsc10 scaled \magstep2
\font\fourteensf=cmss10 scaled \magstep2

\font\seventeenrm=cmr10 scaled \magstep3
\font\seventeenbf=cmbx10 scaled \magstep3
\font\seventeenit=cmti10 scaled \magstep3
\font\seventeeni=cmmi10 scaled \magstep3	\skewchar\seventeeni='177
\font\seventeenmib=cmmib10 scaled \magstep3	\skewchar\seventeenmib='177
\font\seventeensy=cmsy10 scaled \magstep3	\skewchar\seventeensy='60
\font\seventeensyb=cmbsy10 scaled \magstep3	\skewchar\seventeensyb='60
\font\seventeensl=cmsl10 scaled \magstep3
\font\seventeentt=cmtt10 scaled \magstep3	\hyphenchar\seventeentt=-1
\font\seventeencsc=cmcsc10 scaled \magstep3
\font\seventeensf=cmss10 scaled \magstep3

\def\@typeface{Computer Modern} 

\def\hexnumber@#1{\ifnum#1<10 \number#1\else
 \ifnum#1=10 A\else\ifnum#1=11 B\else\ifnum#1=12 C\else
 \ifnum#1=13 D\else\ifnum#1=14 E\else\ifnum#1=15 F\fi\fi\fi\fi\fi\fi\fi}

\def\mib{\hexnumber@\mibfam}
\def\syb{\hexnumber@\sybfam}

\def\makestrut{%
  \setbox\strutbox=\hbox{%
    \vrule height.7\baselineskip depth.3\baselineskip width 0pt}%
}

\def\bls#1{%
  \normalbaselineskip=#1%
  \normalbaselines%
  \makestrut%
}

%

\newfam\mibfam 
\newfam\sybfam 
\newfam\scfam  
\newfam\sffam  

\def\em{\ifdim\fontdimen1\font>0 \rm\else\it\fi}

\textfont3=\tenex
\scriptfont3=\tenex
\scriptscriptfont3=\tenex

\def\eightpoint{
  \def\rm{\fam0\eightrm}%
  \textfont0=\eightrm \scriptfont0=\sevenrm \scriptscriptfont0=\fiverm%
  \textfont1=\eighti  \scriptfont1=\seveni  \scriptscriptfont1=\fivei%
  \textfont2=\eightsy \scriptfont2=\sevensy \scriptscriptfont2=\fivesy%
  \textfont\itfam=\eightit\def\it{\fam\itfam\eightit}%
  \textfont\bffam=\eightbf%
    \scriptfont\bffam=\sevenbf%
      \scriptscriptfont\bffam=\fivebf%
  \def\bf{\fam\bffam\eightbf}%
  \textfont\slfam=\eightsl\def\sl{\fam\slfam\eightsl}%
  \textfont\ttfam=\eighttt\def\tt{\fam\ttfam\eighttt}%
  \textfont\scfam=\eightcsc\def\sc{\fam\scfam\eightcsc}%
  \textfont\sffam=\eightsf\def\sf{\fam\sffam\eightsf}%
  \textfont\mibfam=\eightmib%
  \textfont\sybfam=\eightsyb%
  \bls{10pt}%
}

\def\ninepoint{
  \def\rm{\fam0\ninerm}%
  \textfont0=\ninerm \scriptfont0=\sevenrm \scriptscriptfont0=\fiverm%
  \textfont1=\ninei  \scriptfont1=\seveni  \scriptscriptfont1=\fivei%
  \textfont2=\ninesy \scriptfont2=\sevensy \scriptscriptfont2=\fivesy%
  \textfont\itfam=\nineit\def\it{\fam\itfam\nineit}%
  \textfont\bffam=\ninebf%
    \scriptfont\bffam=\sevenbf%
      \scriptscriptfont\bffam=\fivebf%
  \def\bf{\fam\bffam\ninebf}%
  \textfont\slfam=\ninesl\def\sl{\fam\slfam\ninesl}%
  \textfont\ttfam=\ninett\def\tt{\fam\ttfam\ninett}%
  \textfont\scfam=\ninecsc\def\sc{\fam\scfam\ninecsc}%
  \textfont\sffam=\ninesf\def\sf{\fam\sffam\ninesf}%
  \textfont\mibfam=\ninemib%
  \textfont\sybfam=\ninesyb%
  \bls{12pt}%
}

\def\tenpoint{
  \def\rm{\fam0\tenrm}%
  \textfont0=\tenrm \scriptfont0=\sevenrm \scriptscriptfont0=\fiverm%
  \textfont1=\teni  \scriptfont1=\seveni  \scriptscriptfont1=\fivei%
  \textfont2=\tensy \scriptfont2=\sevensy \scriptscriptfont2=\fivesy%
  \textfont\itfam=\tenit\def\it{\fam\itfam\tenit}%
  \textfont\bffam=\tenbf%
    \scriptfont\bffam=\sevenbf%
      \scriptscriptfont\bffam=\fivebf%
  \def\bf{\fam\bffam\tenbf}%
  \textfont\slfam=\tensl\def\sl{\fam\slfam\tensl}%
  \textfont\ttfam=\tentt\def\tt{\fam\ttfam\tentt}%
  \textfont\scfam=\tencsc\def\sc{\fam\scfam\tencsc}%
  \textfont\sffam=\tensf\def\sf{\fam\sffam\tensf}%
  \textfont\mibfam=\tenmib%
  \textfont\sybfam=\tensyb%
  \bls{12pt}%
}

\def\elevenpoint{
  \def\rm{\fam0\elevenrm}%
  \textfont0=\elevenrm \scriptfont0=\eightrm \scriptscriptfont0=\fiverm%
  \textfont1=\eleveni  \scriptfont1=\eighti  \scriptscriptfont1=\fivei%
  \textfont2=\elevensy \scriptfont2=\eightsy \scriptscriptfont2=\fivesy%
  \textfont\itfam=\elevenit\def\it{\fam\itfam\elevenit}%
  \textfont\bffam=\elevenbf%
    \scriptfont\bffam=\eightbf%
      \scriptscriptfont\bffam=\fivebf%
  \def\bf{\fam\bffam\elevenbf}%
  \textfont\slfam=\elevensl\def\sl{\fam\slfam\elevensl}%
  \textfont\ttfam=\eleventt\def\tt{\fam\ttfam\eleventt}%
  \textfont\scfam=\elevencsc\def\sc{\fam\scfam\elevencsc}%
  \textfont\sffam=\elevensf\def\sf{\fam\sffam\elevensf}%
  \textfont\mibfam=\elevenmib%
  \textfont\sybfam=\elevensyb%
  \bls{13pt}%
}

\def\fourteenpoint{
  \def\rm{\fam0\fourteenrm}%
  \textfont0\fourteenrm  \scriptfont0\tenrm  \scriptscriptfont0\sevenrm%
  \textfont1\fourteeni   \scriptfont1\teni   \scriptscriptfont1\seveni%
  \textfont2\fourteensy  \scriptfont2\tensy  \scriptscriptfont2\sevensy%
  \textfont\itfam=\fourteenit\def\it{\fam\itfam\fourteenit}%
  \textfont\bffam=\fourteenbf%
    \scriptfont\bffam=\tenbf%
      \scriptscriptfont\bffam=\sevenbf%
  \def\bf{\fam\bffam\fourteenbf}%
  \textfont\slfam=\fourteensl\def\sl{\fam\slfam\fourteensl}%
  \textfont\ttfam=\fourteentt\def\tt{\fam\ttfam\fourteentt}%
  \textfont\scfam=\fourteencsc\def\sc{\fam\scfam\fourteencsc}%
  \textfont\sffam=\fourteensf\def\sf{\fam\sffam\fourteensf}%
  \textfont\mibfam=\fourteenmib%
  \textfont\sybfam=\fourteensyb%
  \bls{17pt}%
}

\def\seventeenpoint{
  \def\rm{\fam0\seventeenrm}%
  \textfont0\seventeenrm  \scriptfont0\elevenrm  \scriptscriptfont0\ninerm%
  \textfont1\seventeeni   \scriptfont1\eleveni   \scriptscriptfont1\ninei%
  \textfont2\seventeensy  \scriptfont2\elevensy  \scriptscriptfont2\ninesy%
  \textfont\itfam=\seventeenit\def\it{\fam\itfam\seventeenit}%
  \textfont\bffam=\seventeenbf%
    \scriptfont\bffam=\elevenbf%
      \scriptscriptfont\bffam=\ninebf%
  \def\bf{\fam\bffam\seventeenbf}%
  \textfont\slfam=\seventeensl\def\sl{\fam\slfam\seventeensl}%
  \textfont\ttfam=\seventeentt\def\tt{\fam\ttfam\seventeentt}%
  \textfont\scfam=\seventeencsc\def\sc{\fam\scfam\seventeencsc}%
  \textfont\sffam=\seventeensf\def\sf{\fam\sffam\seventeensf}%
  \textfont\mibfam=\seventeenmib%
  \textfont\sybfam=\seventeensyb%
  \bls{20pt}%
}

\lineskip=1pt      \normallineskip=\lineskip
\lineskiplimit=0pt \normallineskiplimit=\lineskiplimit




\def\Nulle{0}  
\def\Aue{1}    
\def\Afe{2}    
\def\Sue{4}    
\def\Hae{5}    
\def\Hbe{6}    
\def\Hce{7}    
\def\Hde{8}    
\def\Kwe{9}    
\def\Txe{10}   
\def\Lie{11}   
\def\Bbe{12}   


\newdimen\DimenA
\newbox\BoxA

\newcount\LastMac \LastMac=\Nulle
\newcount\HeaderNumber \HeaderNumber=0
\newcount\DefaultHeader \DefaultHeader=\HeaderNumber
\newskip\Indent

\newskip\half      \half=5.5pt plus 1.5pt minus 2.25pt
\newskip\one       \one=11pt plus 3pt minus 5.5pt
\newskip\onehalf   \onehalf=16.5pt plus 5.5pt minus 8.25pt
\newskip\two       \two=22pt plus 5.5pt minus 11pt

\def\Half{\vskip-\lastskip\vskip\half}
\def\One{\vskip-\lastskip\vskip\one}
\def\OneHalf{\vskip-\lastskip\vskip\onehalf}
\def\Two{\vskip-\lastskip\vskip\two}


\def\rTenPT{10pt plus \Feathering}

\def\TenPT{10pt plus \Feathering} 
\def\ElevenPT{11pt plus \Feathering}

\def\Raggedright{
 \rightskip=0pt plus \hsize
}

\def\Fullout{
\rightskip=0pt
}

\def\Hang#1#2{
 \hangindent=#1
 \hangafter=#2
}

\def\EveryMac{
 \Fullout
 \everypar{}
}



\def\title#1{
 \EveryMac
 \LastMac=\Nulle
 \global\HeaderNumber=0
 \global\DefaultHeader=1
 \vbox to 1pc{\vss}
 \seventeenpoint
 \Raggedright
 \noindent \bf #1
}

\def\author#1{
 \EveryMac
 \ifnum\LastMac=\Afe \OneHalf
  \else \Two
 \fi
 \LastMac=\Aue
 \fourteenpoint
 \Raggedright
 \noindent \rm #1\par
 \vskip 3pt\relax
}

\def\affiliation#1{
 \EveryMac
 \LastMac=\Afe
 \eightpoint\bls{\TenPT}
 \Raggedright
 \noindent \it #1\par
}

\def\abstract{%
 \EveryMac
 \Two
 \LastMac=\Sue
 \everypar{\Hang{11pc}{0}}
 \noindent\ninebf ABSTRACT\par
 \tenpoint\bls{\ElevenPT}
 \Fullout
 \noindent\rm
}

\def\keywords{
 \EveryMac
 \Half
 \LastMac=\Kwe
 \everypar{\Hang{11pc}{0}}
 \tenpoint\bls{\ElevenPT}
 \Fullout
 \noindent\hbox{\bf Key words:\ }
 \rm
}


\def\maketitle{%
  \Two%
  \EndOpening%
  \MakePage%
}


\def\pageoffset#1#2{\hoffset=#1\relax\voffset=#2\relax}


\def\Autonumber{
 \global\AutoNumbertrue  
}

\newif\ifAutoNumber \AutoNumberfalse
\newcount\Sec        
\newcount\SecSec
\newcount\SecSecSec

\Sec=0

\def\:{\let\@sptoken= } \:  
\def\:{\@xifnch} \expandafter\def\: {\futurelet\@tempc\@ifnch}

\def\@ifnextchar#1#2#3{%
  \let\@tempMACe #1%
  \def\@tempMACa{#2}%
  \def\@tempMACb{#3}%
  \futurelet \@tempMACc\@ifnch%
}

\def\@ifnch{%
\ifx \@tempMACc \@sptoken%
  \let\@tempMACd\@xifnch%
\else%
  \ifx \@tempMACc \@tempMACe%
    \let\@tempMACd\@tempMACa%
  \else%
    \let\@tempMACd\@tempMACb%
  \fi%
\fi%
\@tempMACd%
}

\def\@ifstar#1#2{\@ifnextchar *{\def\@tempMACa*{#1}\@tempMACa}{#2}}

\def\section{\@ifstar{\@ssection}{\@section}}

\def\@section#1{
 \EveryMac
 \Two
 \LastMac=\Hae
 \ninepoint\bls{\ElevenPT}
 \bf
 \Raggedright
 \ifAutoNumber
  \advance\Sec by 1
  \noindent\number\Sec\hskip 1pc \uppercase{#1}
  \SecSec=0
 \else
  \noindent \uppercase{#1}
 \fi
 \nobreak
}

\def\@ssection#1{
 \EveryMac
 \ifnum\LastMac=\Hae \Half
  \else \OneHalf
 \fi
 \LastMac=\Hae
 \tenpoint\bls{\ElevenPT}
 \bf
 \Raggedright
 \noindent\uppercase{#1}
}

\def\subsection#1{
 \EveryMac
 \ifnum\LastMac=\Hae \Half
  \else \OneHalf
 \fi
 \LastMac=\Hbe
 \tenpoint\bls{\ElevenPT}
 \bf
 \Raggedright
 \ifAutoNumber
  \advance\SecSec by 1
  \noindent\number\Sec.\number\SecSec
  \hskip 1pc #1
  \SecSecSec=0
 \else
  \noindent #1
 \fi
 \nobreak
}

\def\subsubsection#1{
 \EveryMac
 \ifnum\LastMac=\Hbe \Half
  \else \OneHalf
 \fi
 \LastMac=\Hce
 \ninepoint\bls{\ElevenPT}
 \it
 \Raggedright
 \ifAutoNumber
  \advance\SecSecSec by 1
  \noindent\number\Sec.\number\SecSec.\number\SecSecSec
  \hskip 1pc #1
 \else
  \noindent #1
 \fi
 \nobreak
}

\def\paragraph#1{
 \EveryMac
 \One
 \LastMac=\Hde
 \ninepoint\bls{\ElevenPT}
 \noindent \it #1
 \rm
}


\def\tx{
 \EveryMac
 \ifnum\LastMac=\Lie \Half\fi
 \ifnum\LastMac=\Hae \nobreak\Half\fi
 \ifnum\LastMac=\Hbe \nobreak\Half\fi
 \ifnum\LastMac=\Hce \nobreak\Half\fi
 \ifnum\LastMac=\Lie \else \noindent\fi
 \LastMac=\Txe
 \ninepoint\bls{\ElevenPT}
 \rm
}


\def\item{
 \par
 \EveryMac
 \ifnum\LastMac=\Lie
  \else \Half
 \fi
 \LastMac=\Lie
 \ninepoint\bls{\ElevenPT}
 \rm
}


\def\bibitem{
 \par
 \EveryMac
 \ifnum\LastMac=\Bbe
  \else \Half
 \fi
 \LastMac=\Bbe
 \Hang{1.5em}{1}
 \eightpoint\bls{\TenPT}
 \Raggedright
 \noindent \rm
}


\newtoks\CatchLine

\def\@journal{Mon.\ Not.\ R.\ Astron.\ Soc.\ }  
\def\@pubyear{1993}        
\def\@pagerange{000--000}  
\def\@volume{000}          
\def\@microfiche{}         %

\def\pubyear#1{\gdef\@pubyear{#1}\@makecatchline}
\def\pagerange#1{\gdef\@pagerange{#1}\@makecatchline}
\def\volume#1{\gdef\@volume{#1}\@makecatchline}
\def\microfiche#1{\gdef\@microfiche{and Microfiche\ #1}\@makecatchline}

\def\@makecatchline{%
  \global\CatchLine{%
    {\rm \@journal {\bf \@volume},\ \@pagerange\ (\@pubyear)\ \@microfiche}}%
}

\@makecatchline 

\newtoks\LeftHeader
\def\shortauthor#1{
 \global\LeftHeader{#1}
}

\newtoks\RightHeader
\def\shorttitle#1{
 \global\RightHeader{#1}
}

\def\PageHead{
 \EveryMac
 \ifnum\HeaderNumber=1 \Pagehead
  \else \Catchline
 \fi
}

\def\Catchline{%
 \vbox to 0pt{\vskip-22.5pt
  \hbox to \PageWidth{\vbox to8.5pt{}\noindent
  \eightpoint\the\CatchLine\hfill}\vss}
 \nointerlineskip
}

\def\Pagehead{%
 \ifodd\pageno
   \vbox to 0pt{\vskip-22.5pt
   \hbox to \PageWidth{\vbox to8.5pt{}\elevenpoint\it\noindent
    \hfill\the\RightHeader\hskip1.5em\rm\folio}\vss}
 \else
   \vbox to 0pt{\vskip-22.5pt
   \hbox to \PageWidth{\vbox to8.5pt{}\elevenpoint\rm\noindent
   \folio\hskip1.5em\it\the\LeftHeader\hfill}\vss}
 \fi
 \nointerlineskip
}

\def\PageFoot{} 

\def\authorcomment#1{%
  \gdef\PageFoot{%
    \nointerlineskip%
    \vbox to 22pt{\vfil%
      \hbox to \PageWidth{\elevenpoint\rm\noindent \hfil #1 \hfil}}%
  }%
}

\everydisplay{\displaysetup}

\newif\ifeqno
\newif\ifleqno

\def\displaysetup#1$${%
 \displaytest#1\eqno\eqno\displaytest
}

\def\displaytest#1\eqno#2\eqno#3\displaytest{%
 \if!#3!\ldisplaytest#1\leqno\leqno\ldisplaytest
 \else\eqnotrue\leqnofalse\def\eqn{#2}\def\eq{#1}\fi
 \generaldisplay$$}

\def\ldisplaytest#1\leqno#2\leqno#3\ldisplaytest{%
 \def\eq{#1}%
 \if!#3!\eqnofalse\else\eqnotrue\leqnotrue
  \def\eqn{#2}\fi}

\def\generaldisplay{%
\ifeqno \ifleqno 
   \hbox to \hsize{\noindent
     $\displaystyle\eq$\hfil$\displaystyle\eqn$}
  \else
    \hbox to \hsize{\noindent
     $\displaystyle\eq$\hfil$\displaystyle\eqn$}
  \fi
 \else
 \hbox to \hsize{\vbox{\noindent
  $\displaystyle\eq$\hfil}}
 \fi
}

\def\@notice{%
  \par\Two%
  \bls{12pt}%
  \noindent\tenrm This paper has been produced using the Blackwell
                  Scientific Publications \TeX\ macros.%
}

\outer\def\bye{\@notice\par\vfill\supereject\end}

\everyjob{%
  \Warn{Monthly notices of the RAS journal style (\@typeface)\space
        v\@version,\space \@verdate.}\Warn{}%
}




\newif\if@debug \@debugfalse  

\def\Print#1{\if@debug\immediate\write16{#1}\else \fi}
\def\Warn#1{\immediate\write16{#1}}
\def\wlog#1{}

\newcount\Iteration 

\newif\ifFigureBoxes  
\FigureBoxestrue

\def\Single{0} \def\Double{1}                 
\def\Figure{0} \def\Table{1}                  

\def\InStack{0}  
\def\InZoneA{1}
\def\InZoneB{2}
\def\InZoneC{3}

\newcount\TEMPCOUNT 
\newdimen\TEMPDIMEN 
\newbox\TEMPBOX     
\newbox\VOIDBOX     

\newcount\LengthOfStack 
\newcount\MaxItems      
\newcount\StackPointer
\newcount\Point         
\newcount\NextFigure    
\newcount\NextTable     
\newcount\NextItem      

\newcount\StatusStack   
\newcount\NumStack      
\newcount\TypeStack     
\newcount\SpanStack     
\newcount\BoxStack      

\newcount\ItemSTATUS    
\newcount\ItemNUMBER    
\newcount\ItemTYPE      
\newcount\ItemSPAN      
\newbox\ItemBOX         
\newdimen\ItemSIZE      

\newdimen\PageHeight    
\newdimen\TextLeading   
\newdimen\Feathering    
\newcount\LinesPerPage  
\newdimen\ColumnWidth   
\newdimen\ColumnGap     
\newdimen\PageWidth     
\newdimen\BodgeHeight   
\newcount\Leading       

\newdimen\ZoneBSize  
\newdimen\TextSize   
\newbox\ZoneABOX     
\newbox\ZoneBBOX     
\newbox\ZoneCBOX     

\newif\ifFirstSingleItem
\newif\ifFirstZoneA
\newif\ifMakePageInComplete
\newif\ifMoreFigures \MoreFiguresfalse 
\newif\ifMoreTables  \MoreTablesfalse  

\newif\ifFigInZoneB 
\newif\ifFigInZoneC 
\newif\ifTabInZoneB 
\newif\ifTabInZoneC

\newif\ifZoneAFullPage

\newbox\MidBOX    
\newbox\LeftBOX
\newbox\RightBOX
\newbox\PageBOX   

\newif\ifLeftCOL  
\LeftCOLtrue

\newdimen\ZoneBAdjust

\newcount\ItemFits
\def\Yes{1}
\def\No{2}




\MaxItems=15
\NextFigure=0        
\NextTable=1

\BodgeHeight=6pt
\TextLeading=11pt    
\Leading=11
\Feathering=0pt      
\LinesPerPage=61     
\topskip=\TextLeading
\ColumnWidth=20pc    
\ColumnGap=2pc       

\def\ItemSep{\vskip \TextLeading plus \TextLeading minus 4pt}

\FigureBoxesfalse 

\parskip=0pt
\parindent=18pt
\widowpenalty=0
\clubpenalty=10000
\tolerance=1500
\hbadness=1500
\abovedisplayskip=6pt plus 2pt minus 2pt
\belowdisplayskip=6pt plus 2pt minus 2pt
\abovedisplayshortskip=6pt plus 2pt minus 2pt
\belowdisplayshortskip=6pt plus 2pt minus 2pt

\PageHeight=\TextLeading 
\multiply\PageHeight by \LinesPerPage
\advance\PageHeight by \topskip

\PageWidth=2\ColumnWidth
\advance\PageWidth by \ColumnGap




\newcount\DUMMY \StatusStack=\allocationnumber
\newcount\DUMMY \newcount\DUMMY \newcount\DUMMY 
\newcount\DUMMY \newcount\DUMMY \newcount\DUMMY 
\newcount\DUMMY \newcount\DUMMY \newcount\DUMMY
\newcount\DUMMY \newcount\DUMMY \newcount\DUMMY 
\newcount\DUMMY \newcount\DUMMY \newcount\DUMMY

\newcount\DUMMY \NumStack=\allocationnumber
\newcount\DUMMY \newcount\DUMMY \newcount\DUMMY 
\newcount\DUMMY \newcount\DUMMY \newcount\DUMMY 
\newcount\DUMMY \newcount\DUMMY \newcount\DUMMY 
\newcount\DUMMY \newcount\DUMMY \newcount\DUMMY 
\newcount\DUMMY \newcount\DUMMY \newcount\DUMMY

\newcount\DUMMY \TypeStack=\allocationnumber
\newcount\DUMMY \newcount\DUMMY \newcount\DUMMY 
\newcount\DUMMY \newcount\DUMMY \newcount\DUMMY 
\newcount\DUMMY \newcount\DUMMY \newcount\DUMMY 
\newcount\DUMMY \newcount\DUMMY \newcount\DUMMY 
\newcount\DUMMY \newcount\DUMMY \newcount\DUMMY

\newcount\DUMMY \SpanStack=\allocationnumber
\newcount\DUMMY \newcount\DUMMY \newcount\DUMMY 
\newcount\DUMMY \newcount\DUMMY \newcount\DUMMY 
\newcount\DUMMY \newcount\DUMMY \newcount\DUMMY 
\newcount\DUMMY \newcount\DUMMY \newcount\DUMMY 
\newcount\DUMMY \newcount\DUMMY \newcount\DUMMY

\newbox\DUMMY   \BoxStack=\allocationnumber
\newbox\DUMMY   \newbox\DUMMY \newbox\DUMMY 
\newbox\DUMMY   \newbox\DUMMY \newbox\DUMMY 
\newbox\DUMMY   \newbox\DUMMY \newbox\DUMMY 
\newbox\DUMMY   \newbox\DUMMY \newbox\DUMMY 
\newbox\DUMMY   \newbox\DUMMY \newbox\DUMMY

\def\wlog{\immediate\write-1}


\def\GetItemAll#1{%
 \GetItemSTATUS{#1}
 \GetItemNUMBER{#1}
 \GetItemTYPE{#1}
 \GetItemSPAN{#1}
 \GetItemBOX{#1}
}

\def\GetItemSTATUS#1{%
 \Point=\StatusStack
 \advance\Point by #1
 \global\ItemSTATUS=\count\Point
}

\def\GetItemNUMBER#1{%
 \Point=\NumStack
 \advance\Point by #1
 \global\ItemNUMBER=\count\Point
}

\def\GetItemTYPE#1{%
 \Point=\TypeStack
 \advance\Point by #1
 \global\ItemTYPE=\count\Point
}

\def\GetItemSPAN#1{%
 \Point\SpanStack
 \advance\Point by #1
 \global\ItemSPAN=\count\Point
}

\def\GetItemBOX#1{%
 \Point=\BoxStack
 \advance\Point by #1
 \global\setbox\ItemBOX=\vbox{\copy\Point}
 \global\ItemSIZE=\ht\ItemBOX
 \global\advance\ItemSIZE by \dp\ItemBOX
 \TEMPCOUNT=\ItemSIZE
 \divide\TEMPCOUNT by \Leading
 \divide\TEMPCOUNT by 65536
 \advance\TEMPCOUNT by 1
 \ItemSIZE=\TEMPCOUNT pt
 \global\multiply\ItemSIZE by \Leading
}


\def\JoinStack{%
 \ifnum\LengthOfStack=\MaxItems 
  \Warn{WARNING: Stack is full...some items will be lost!}
 \else
  \Point=\StatusStack
  \advance\Point by \LengthOfStack
  \global\count\Point=\ItemSTATUS
  \Point=\NumStack
  \advance\Point by \LengthOfStack
  \global\count\Point=\ItemNUMBER
  \Point=\TypeStack
  \advance\Point by \LengthOfStack
  \global\count\Point=\ItemTYPE
  \Point\SpanStack
  \advance\Point by \LengthOfStack
  \global\count\Point=\ItemSPAN
  \Point=\BoxStack
  \advance\Point by \LengthOfStack
  \global\setbox\Point=\vbox{\copy\ItemBOX}
  \global\advance\LengthOfStack by 1
  \ifnum\ItemTYPE=\Figure 
   \global\MoreFigurestrue
  \else
   \global\MoreTablestrue
  \fi
 \fi
}


\def\LeaveStack#1{%
 {\Iteration=#1
 \loop
 \ifnum\Iteration<\LengthOfStack
  \advance\Iteration by 1
  \GetItemSTATUS{\Iteration}
   \advance\Point by -1
   \global\count\Point=\ItemSTATUS
  \GetItemNUMBER{\Iteration}
   \advance\Point by -1
   \global\count\Point=\ItemNUMBER
  \GetItemTYPE{\Iteration}
   \advance\Point by -1
   \global\count\Point=\ItemTYPE
  \GetItemSPAN{\Iteration}
   \advance\Point by -1
   \global\count\Point=\ItemSPAN
  \GetItemBOX{\Iteration}
   \advance\Point by -1
   \global\setbox\Point=\vbox{\copy\ItemBOX}
 \repeat}
 \global\advance\LengthOfStack by -1
}


\newif\ifStackNotClean

\def\CleanStack{%
 \StackNotCleantrue
 {\Iteration=0
  \loop
   \ifStackNotClean
    \GetItemSTATUS{\Iteration}
    \ifnum\ItemSTATUS=\InStack
     \advance\Iteration by 1
     \else
      \LeaveStack{\Iteration}
    \fi
   \ifnum\LengthOfStack<\Iteration
    \StackNotCleanfalse
   \fi
 \repeat}
}


\def\FindItem#1#2{%
 \global\StackPointer=-1 
 {\Iteration=0
  \loop
  \ifnum\Iteration<\LengthOfStack
   \GetItemSTATUS{\Iteration}
   \ifnum\ItemSTATUS=\InStack
    \GetItemTYPE{\Iteration}
    \ifnum\ItemTYPE=#1
     \GetItemNUMBER{\Iteration}
     \ifnum\ItemNUMBER=#2
      \global\StackPointer=\Iteration
      \Iteration=\LengthOfStack 
     \fi
    \fi
   \fi
  \advance\Iteration by 1
 \repeat}
}


\def\FindNext{%
 \global\StackPointer=-1 
 {\Iteration=0
  \loop
  \ifnum\Iteration<\LengthOfStack
   \GetItemSTATUS{\Iteration}
   \ifnum\ItemSTATUS=\InStack
    \GetItemTYPE{\Iteration}
   \ifnum\ItemTYPE=\Figure
    \ifMoreFigures
      \global\NextItem=\Figure
      \global\StackPointer=\Iteration
      \Iteration=\LengthOfStack 
    \fi
   \fi
   \ifnum\ItemTYPE=\Table
    \ifMoreTables
      \global\NextItem=\Table
      \global\StackPointer=\Iteration
      \Iteration=\LengthOfStack 
    \fi
   \fi
  \fi
  \advance\Iteration by 1
 \repeat}
}


\def\ChangeStatus#1#2{%
 \Point=\StatusStack
 \advance\Point by #1
 \global\count\Point=#2
}



\def\Zone{\InZoneA}

\ZoneBAdjust=0pt

\def\MakePage{
 \global\ZoneBSize=\PageHeight
 \global\TextSize=\ZoneBSize
 \global\ZoneAFullPagefalse
 \global\topskip=\TextLeading
 \MakePageInCompletetrue
 \MoreFigurestrue
 \MoreTablestrue
 \FigInZoneBfalse
 \FigInZoneCfalse
 \TabInZoneBfalse
 \TabInZoneCfalse
 \global\FirstSingleItemtrue
 \global\FirstZoneAtrue
 \global\setbox\ZoneABOX=\box\VOIDBOX
 \global\setbox\ZoneBBOX=\box\VOIDBOX
 \global\setbox\ZoneCBOX=\box\VOIDBOX
 \loop
  \ifMakePageInComplete
 \FindNext
 \ifnum\StackPointer=-1
  \NextItem=-1
  \MoreFiguresfalse
  \MoreTablesfalse
 \fi
 \ifnum\NextItem=\Figure
   \FindItem{\Figure}{\NextFigure}
   \ifnum\StackPointer=-1 \global\MoreFiguresfalse
   \else
    \GetItemSPAN{\StackPointer}
    \ifnum\ItemSPAN=\Single \def\Zone{\InZoneB}\relax
     \ifFigInZoneC \global\MoreFiguresfalse\fi
    \else
     \def\Zone{\InZoneA}
     \ifFigInZoneB \def\Zone{\InZoneC}\fi
    \fi
   \fi
   \ifMoreFigures\Print{}\FigureItems\fi
 \fi
\ifnum\NextItem=\Table
   \FindItem{\Table}{\NextTable}
   \ifnum\StackPointer=-1 \global\MoreTablesfalse
   \else
    \GetItemSPAN{\StackPointer}
    \ifnum\ItemSPAN=\Single\relax
     \ifTabInZoneC \global\MoreTablesfalse\fi
    \else
     \def\Zone{\InZoneA}
     \ifTabInZoneB \def\Zone{\InZoneC}\fi
    \fi
   \fi
   \ifMoreTables\Print{}\TableItems\fi
 \fi
   \MakePageInCompletefalse 
   \ifMoreFigures\MakePageInCompletetrue\fi
   \ifMoreTables\MakePageInCompletetrue\fi
 \repeat
 \ifZoneAFullPage
  \global\TextSize=0pt
  \global\ZoneBSize=0pt
  \global\vsize=0pt\relax
  \global\topskip=0pt\relax
  \vbox to 0pt{\vss}
  \eject
 \else
 \global\advance\ZoneBSize by -\ZoneBAdjust
 \global\vsize=\ZoneBSize
 \global\hsize=\ColumnWidth
 \global\ZoneBAdjust=0pt
 \ifdim\TextSize<23pt
 \Warn{}
 \Warn{* Making column fall short: TextSize=\the\TextSize *}
 \vskip-\lastskip\eject\fi
 \fi
}

\def\MakeRightCol{
 \global\TextSize=\ZoneBSize
 \MakePageInCompletetrue
 \MoreFigurestrue
 \MoreTablestrue
 \global\FirstSingleItemtrue
 \global\setbox\ZoneBBOX=\box\VOIDBOX
 \def\Zone{\InZoneB}
 \loop
  \ifMakePageInComplete
 \FindNext
 \ifnum\StackPointer=-1
  \NextItem=-1
  \MoreFiguresfalse
  \MoreTablesfalse
 \fi
 \ifnum\NextItem=\Figure
   \FindItem{\Figure}{\NextFigure}
   \ifnum\StackPointer=-1 \MoreFiguresfalse
   \else
    \GetItemSPAN{\StackPointer}
    \ifnum\ItemSPAN=\Double\relax
     \MoreFiguresfalse\fi
   \fi
   \ifMoreFigures\Print{}\FigureItems\fi
 \fi
 \ifnum\NextItem=\Table
   \FindItem{\Table}{\NextTable}
   \ifnum\StackPointer=-1 \MoreTablesfalse
   \else
    \GetItemSPAN{\StackPointer}
    \ifnum\ItemSPAN=\Double\relax
     \MoreTablesfalse\fi
   \fi
   \ifMoreTables\Print{}\TableItems\fi
 \fi
   \MakePageInCompletefalse 
   \ifMoreFigures\MakePageInCompletetrue\fi
   \ifMoreTables\MakePageInCompletetrue\fi
 \repeat
 \ifZoneAFullPage
  \global\TextSize=0pt
  \global\ZoneBSize=0pt
  \global\vsize=0pt\relax
  \global\topskip=0pt\relax
  \vbox to 0pt{\vss}
  \eject
 \else
 \global\vsize=\ZoneBSize
 \global\hsize=\ColumnWidth
 \ifdim\TextSize<23pt
 \Warn{}
 \Warn{* Making column fall short: TextSize=\the\TextSize *}
 \vskip-\lastskip\eject\fi
\fi
}

\def\FigureItems{
 \Print{Considering...}
 \ShowItem{\StackPointer}
 \GetItemBOX{\StackPointer} 
 \GetItemSPAN{\StackPointer}
  \CheckFitInZone 
  \ifnum\ItemFits=\Yes
   \ifnum\ItemSPAN=\Single
     \ChangeStatus{\StackPointer}{\InZoneB} 
     \global\FigInZoneBtrue
     \ifFirstSingleItem
      \hbox{}\vskip-\BodgeHeight
     \global\advance\ItemSIZE by \TextLeading
     \fi
     \unvbox\ItemBOX\ItemSep
     \global\FirstSingleItemfalse
     \global\advance\TextSize by -\ItemSIZE
     \global\advance\TextSize by -\TextLeading
   \else
    \ifFirstZoneA
     \global\advance\ItemSIZE by \TextLeading
     \global\FirstZoneAfalse\fi
    \global\advance\TextSize by -\ItemSIZE
    \global\advance\TextSize by -\TextLeading
    \global\advance\ZoneBSize by -\ItemSIZE
    \global\advance\ZoneBSize by -\TextLeading
    \ifFigInZoneB\relax
     \else
     \ifdim\TextSize<3\TextLeading
     \global\ZoneAFullPagetrue
     \fi
    \fi
    \ChangeStatus{\StackPointer}{\Zone}
    \ifnum\Zone=\InZoneC \global\FigInZoneCtrue\fi
  \fi
   \Print{TextSize=\the\TextSize}
   \Print{ZoneBSize=\the\ZoneBSize}
  \global\advance\NextFigure by 1
   \Print{This figure has been placed.}
  \else
   \Print{No space available for this figure...holding over.}
   \Print{}
   \global\MoreFiguresfalse
  \fi
}

\def\TableItems{
 \Print{Considering...}
 \ShowItem{\StackPointer}
 \GetItemBOX{\StackPointer} 
 \GetItemSPAN{\StackPointer}
  \CheckFitInZone 
  \ifnum\ItemFits=\Yes
   \ifnum\ItemSPAN=\Single
    \ChangeStatus{\StackPointer}{\InZoneB}
     \global\TabInZoneBtrue
     \ifFirstSingleItem
      \hbox{}\vskip-\BodgeHeight
     \global\advance\ItemSIZE by \TextLeading
     \fi
     \unvbox\ItemBOX\ItemSep
     \global\FirstSingleItemfalse
     \global\advance\TextSize by -\ItemSIZE
     \global\advance\TextSize by -\TextLeading
   \else
    \ifFirstZoneA
    \global\advance\ItemSIZE by \TextLeading
    \global\FirstZoneAfalse\fi
    \global\advance\TextSize by -\ItemSIZE
    \global\advance\TextSize by -\TextLeading
    \global\advance\ZoneBSize by -\ItemSIZE
    \global\advance\ZoneBSize by -\TextLeading
    \ifFigInZoneB\relax
     \else
     \ifdim\TextSize<3\TextLeading
     \global\ZoneAFullPagetrue
     \fi
    \fi
    \ChangeStatus{\StackPointer}{\Zone}
    \ifnum\Zone=\InZoneC \global\TabInZoneCtrue\fi
   \fi
  \global\advance\NextTable by 1
   \Print{This table has been placed.}
  \else
  \Print{No space available for this table...holding over.}
   \Print{}
   \global\MoreTablesfalse
  \fi
}


\def\CheckFitInZone{%
{\advance\TextSize by -\ItemSIZE
 \advance\TextSize by -\TextLeading
 \ifFirstSingleItem
  \advance\TextSize by \TextLeading
 \fi
 \ifnum\Zone=\InZoneA\relax
  \else \advance\TextSize by -\ZoneBAdjust
 \fi
 \ifdim\TextSize<3\TextLeading \global\ItemFits=\No
 \else \global\ItemFits=\Yes\fi}
}

\def\BF#1#2{
 \ItemSTATUS=\InStack
 \ItemNUMBER=#1
 \ItemTYPE=\Figure
 \if#2S \ItemSPAN=\Single
  \else \ItemSPAN=\Double
 \fi
 \setbox\ItemBOX=\vbox{}
}

\def\BT#1#2{
 \ItemSTATUS=\InStack
 \ItemNUMBER=#1
 \ItemTYPE=\Table
 \if#2S \ItemSPAN=\Single
  \else \ItemSPAN=\Double
 \fi
 \setbox\ItemBOX=\vbox{}
}

\def\BeginOpening{%
 \hsize=\PageWidth
 \global\setbox\ItemBOX=\vbox\bgroup
}

\let\begintopmatter=\BeginOpening  

\def\EndOpening{%
 \egroup
 \ItemNUMBER=0
 \ItemTYPE=\Figure
 \ItemSPAN=\Double
 \ItemSTATUS=\InStack
 \JoinStack
}


\newbox\tmpbox

\def\FC#1#2#3#4{%
  \ItemSTATUS=\InStack
  \ItemNUMBER=#1
  \ItemTYPE=\Figure
  \if#2S
    \ItemSPAN=\Single \TEMPDIMEN=\ColumnWidth
  \else
    \ItemSPAN=\Double \TEMPDIMEN=\PageWidth
  \fi
  {\hsize=\TEMPDIMEN
   \global\setbox\ItemBOX=\vbox{%
     \ifFigureBoxes
       \B{\TEMPDIMEN}{#3}
     \else
       \vbox to #3{\vfil}%
     \fi%
     \eightpoint\rm\bls{\rTenPT}%
     \vskip 5.5pt plus 6pt%
     \setbox\tmpbox=\vbox{#4\par}%
     \ifdim\ht\tmpbox>10pt 
       \noindent #4\par%
     \else
       \hbox to \hsize{\hfil #4\hfil}%
     \fi%
   }%
  }%
  \JoinStack%
  \Print{Processing source for figure {\the\ItemNUMBER}}%
}

\let\figure=\FC  

\def\TH#1#2#3#4{%
 \ItemSTATUS=\InStack
 \ItemNUMBER=#1
 \ItemTYPE=\Table
 \if#2S \ItemSPAN=\Single \TEMPDIMEN=\ColumnWidth
  \else \ItemSPAN=\Double \TEMPDIMEN=\PageWidth
 \fi
{\hsize=\TEMPDIMEN
\eightpoint\bls{\rTenPT}\rm
\global\setbox\ItemBOX=\vbox{\noindent#3\vskip 5.5pt plus5.5pt\noindent#4}}
 \JoinStack
 \Print{Processing source for table {\the\ItemNUMBER}}
}

\let\table=\TH  

\def\UnloadZoneA{%
\FirstZoneAtrue
 \Iteration=0
  \loop
   \ifnum\Iteration<\LengthOfStack
    \GetItemSTATUS{\Iteration}
    \ifnum\ItemSTATUS=\InZoneA
     \GetItemBOX{\Iteration}
     \ifFirstZoneA \vbox to \BodgeHeight{\vfil}%
     \FirstZoneAfalse\fi
     \unvbox\ItemBOX\ItemSep
     \LeaveStack{\Iteration}
     \else
     \advance\Iteration by 1
   \fi
 \repeat
}

\def\UnloadZoneC{%
\Iteration=0
  \loop
   \ifnum\Iteration<\LengthOfStack
    \GetItemSTATUS{\Iteration}
    \ifnum\ItemSTATUS=\InZoneC
     \GetItemBOX{\Iteration}
     \ItemSep\unvbox\ItemBOX
     \LeaveStack{\Iteration}
     \else
     \advance\Iteration by 1
   \fi
 \repeat
}


\def\ShowItem#1{
  {\GetItemAll{#1}
  \Print{\the#1:
  {TYPE=\ifnum\ItemTYPE=\Figure Figure\else Table\fi}
  {NUMBER=\the\ItemNUMBER}
  {SPAN=\ifnum\ItemSPAN=\Single Single\else Double\fi}
  {SIZE=\the\ItemSIZE}}}
}

\def\ShowStack{%
 \Print{}
 \Print{LengthOfStack = \the\LengthOfStack}
 \ifnum\LengthOfStack=0 \Print{Stack is empty}\fi
 \Iteration=0
 \loop
 \ifnum\Iteration<\LengthOfStack
  \ShowItem{\Iteration}
  \advance\Iteration by 1
 \repeat
}

\def\B#1#2{%
\hbox{\vrule\kern-0.4pt\vbox to #2{%
\hrule width #1\vfill\hrule}\kern-0.4pt\vrule}
}

\def\Ref#1{\begingroup\global\setbox\TEMPBOX=\vbox{\hsize=2in\noindent#1}\endgroup
\ht1=0pt\dp1=0pt\wd1=0pt\vadjust{\vtop to 0pt{\advance
\hsize0.5pc\kern-10pt\moveright\hsize\box\TEMPBOX\vss}}}

\def\MarkRef#1{\leavevmode\thinspace\hbox{\vrule\vtop
{\vbox{\hrule\kern1pt\hbox{\vphantom{\rm/}\thinspace{\rm#1}%
\thinspace}}\kern1pt\hrule}\vrule}\thinspace}%


\output{%
 \ifLeftCOL
  \global\setbox\LeftBOX=\vbox to \ZoneBSize{\box255\unvbox\ZoneBBOX}
  \global\LeftCOLfalse
  \MakeRightCol
 \else
  \setbox\RightBOX=\vbox to \ZoneBSize{\box255\unvbox\ZoneBBOX}
  \setbox\MidBOX=\hbox{\box\LeftBOX\hskip\ColumnGap\box\RightBOX}
  \setbox\PageBOX=\vbox to \PageHeight{%
  \UnloadZoneA\box\MidBOX\UnloadZoneC}
  \shipout\vbox{\PageHead\box\PageBOX\PageFoot}
  \global\advance\pageno by 1
  \global\HeaderNumber=\DefaultHeader
  \global\LeftCOLtrue
  \CleanStack
  \MakePage
 \fi
}


\catcode `\@=12 
\def\210{{\rm 2--10\,keV}}
\def\ergs{${\rm erg\,s^{-1}}$}
\def\ergcms{${\rm erg\,cm^{-2}\,s^{-1}}$}

\pageoffset{-2pc}{0pc}
\Autonumber
\begintopmatter
\title{{\it ASCA} observations of Deep ROSAT fields - II.  
The 2--10$\,$keV AGN luminosity function}
\author{B.J.Boyle$^1$, I.Georgantopoulos$^2$, A.J.Blair$^2$, G.C.Stewart$^2$, R.E.Griffiths$^3$, T.Shanks$^4$, K.F.Gunn$^4$, O.Almaini$^5$}
\affiliation{$^1$ Anglo-Australian Observatory, PO Box 296, Epping, 
NSW 2121, Australia}
\affiliation{$^2$ Department of Physics \& Astronomy, The University 
of Leicester, Leicester LE1 7RH}
\affiliation{$^3$ Department of Physics, Carnegie Mellon University, 
Wean Hall, 5000 Forbes Ave., Pittsburgh, PA 15213, USA}
\affiliation{$^4$ Physics Department, University of Durham, South Road,
Durham DH1 3LE}
\affiliation{$^5$ Institute of Astronomy, Madingley Road, Cambridge
CB3 0HA}

\shortauthor{B.J.Boyle et al.}
\shorttitle{{\it ASCA} - II: AGN LF}

\abstract 

We present additional optical spectroscopic identifications of sources
identified in three deep {\it ASCA} GIS fields which also form part of
a deep {\it ROSAT} survey.  In total, 26 {\it ASCA} sources have been
detected down to a 2-10$\,$keV flux limit of $S_{\rm 2-10} = 5 \times
10^{-14}\,$\ergcms. LDSS observations have increased the spectroscopic
completeness of the survey to between 65 and 85 per cent, with
identifications for up to 13 QSOs with broad emission lines and 6
objects with narrow emission lines.  Combining these objects with the
AGN identified in the {\it HEAO 1} sample by Grossan, we find
evidence for significant cosmological evolution in the 2-10$\,$keV
band ($\langle{V_e\over V_a}\rangle=0.63\pm 0.03$), fit by a pure
luminosity evolution model; $L \propto (1+z)^k$,
$k=2.04^{+0.16}_{-0.22}$.  The present-epoch 2-10$\,$keV AGN X-ray
luminosity function, $\Phi(L_X)$, is best represented by a two power
law function: $\Phi (L_X) \propto L^{-3.0}$,
$L^*>10^{44.1}\,$erg$\,$s$^{-1}$; $\Phi (L_X) \propto L^{-1.9}$,
$L^*<10^{44.1}\,$erg$\,$s$^{-1}$.  Depending on the extent to which we
extrapolate the $z=0$ AGN luminosity function, we predict a total
contribution to the 2-10$\,$keV X-ray background from QSOs and
narrow-emission-line galaxies which ranges from 48 per cent (for AGN
with $L_X>10^{42}\,$erg$\,$s$^{-1}$) to 80 per cent
($L_X>10^{39}\,$erg$\,$s$^{-1}$).

\keywords X-rays: general -- galaxies: active -- quasars: general
\maketitle
 
\section{Introduction}\tx

The {\it ROSAT} mission has provided us with a significant advance
in our knowledge regarding the nature of the faint X-ray source
population at soft 0.5-2$\,$keV energies.  Deep {\it ROSAT} 
pointings have resolved over half of the soft X-ray background 
(Hasinger et al.\ 1993) and numerous spectroscopic studies of the 
optical counterparts to these sources (Shanks et al.\ 1991,
Carballo et al.\ 1995, Georgantopoulos et al.\ 1996, Bower et al.\
1996) have shown that the sources are predominantly QSOs and 
narrow-emission-line galaxies.

In contrast, until recently, little has been known about the
nature of the X-ray source population at harder energies ($>2\,$keV), 
where the bulk of the energy in the X-ray background (XRB) lies
(see Fabian and Barcons 1992).  
The {\it HEAO 1} A-1 experiment (Wood et al.\ 1984) resolved less than 
5 per cent of the 2-10$\,$keV X-ray background at its
flux limit $S_{\rm 2-10} \sim 5 \times 10^{-12}\,$\ergcms.
 
Data obtained with the {\it ASCA} satellite now allows us to extend
this limiting flux by a factor of almost 100, affording a new insight
into the nature of the faint X-ray source population in the
2-10$\,$keV energy range.  In a previous paper in this series
(Georgantopoulos et al.\ 1997, hereinafter Paper I), we reported on
{\it ASCA} observations of three fields we had previously studied with
{\it ROSAT}.  The limiting flux for the {\it ASCA} survey was $S_{\rm
2-10} \sim 5 \times 10^{-14}\,$\ergcms, resolving approximately 30 per
cent of the 2-10$\,$keV background.  By cross-correlating the
positions of the {\it ASCA} sources with the {\it ROSAT} sources, we
were able to derive optical identifications for approximately half of
the {\it ASCA} sources.  However, there were too few identifications
to yield a definitive answer as to the nature of the majority of the
objects in the {\it ASCA} source list.  We have therefore attempted to
obtain further spectroscopic identifications for the optical
counterparts in our {\it ASCA} source list in order to obtain a
clearer picture of the nature of the faint 2-10$\,$keV X-ray source
population.  In section 2 we describe our optical and X-ray
observations and present an analysis of the source populations
identified in section 3. We discuss in section 4 the implications for
the 2-10$\,$keV background and we present our conclusions in section
5.  Values of $H_0=50\,{\rm km\,s^{-1}\,Mpc^{-1}}$ and $q_0=0.5$ are
assumed throughout this paper.  All co-ordinates are given in equinox
J2000.0.

\vfill\eject

\section{Observations}\tx

\subsection{X-ray}\tx

Full details of the X-ray observations for the three {\it ASCA} GIS
fields used in this study are reported in Paper I.  Briefly, three
fields; GSGP4 (00$^{\rm h}$57$^{m}$25.2$^{s}$
$-$27$^{\circ}$37$'$48$''$), QSF3 (03$^{\rm h}$41$^{m}$44.4$^{s}$
$-$44$^{\circ}$07$'$05$''$), F855 (10$^{\rm
h}$46$^{m}$24.0$^{s}$\break $-$00$^{\circ}$20$'$38$''$), in the deep
{\it ROSAT} survey of Shanks et al. (in preparation) were observed
with the {\it ASCA} GIS for total exposure times up to 218 ksec.  We
detected 26 sources down to a $4\sigma$ threshold.  A total of
$1\,$deg$^{2}$ was searched to a mean flux limit of $S_{\rm 2-10}
\sim 5 \times 10^{-14}\,$\ergcms.

Although the Point Spread Function (PSF) of the {\it ASCA} GIS leads
to relatively poor positional accuracy ($1\sigma$ error $\sim
1\,$arcmin) for the detected sources, we were able to cross-correlate
the {\it ASCA} source catalogue on these three fields with our deep
$4\sigma$ {\it ROSAT} source list on the same fields.  The typical
$1\sigma$ positional error in the {\it ROSAT} positions is
$10\,$arcsec.  The cross-correlation yielded a strong signal between
the positions of detected {\it ASCA} sources and {\it ROSAT} $4\sigma$
source list (see Table 2 in Paper I).  At angular separations less
than 90 arcsec we found 26 matches between the {\it ASCA} and {\it
ROSAT} source positions, compared to 8.5 expected on average.
Three {\it ASCA} sources had more than one {\it ROSAT} counterpart
within 90 arcsec and only three {\it ASCA} sources had no
corresponding {\it ROSAT} counterpart.

Although there will be a number of chance coincidences amongst these
matches, the {\it ASCA}--{\it ROSAT} positional cross-correlation does
yield a much more accurate position for the vast majority of the {\it
ASCA} sources.  Moreover, statistical account can be made for the
number of chance coincidences from the cross-correlation analysis (see
below).

\figure{1}{D}{0mm}{
\psfig{figure=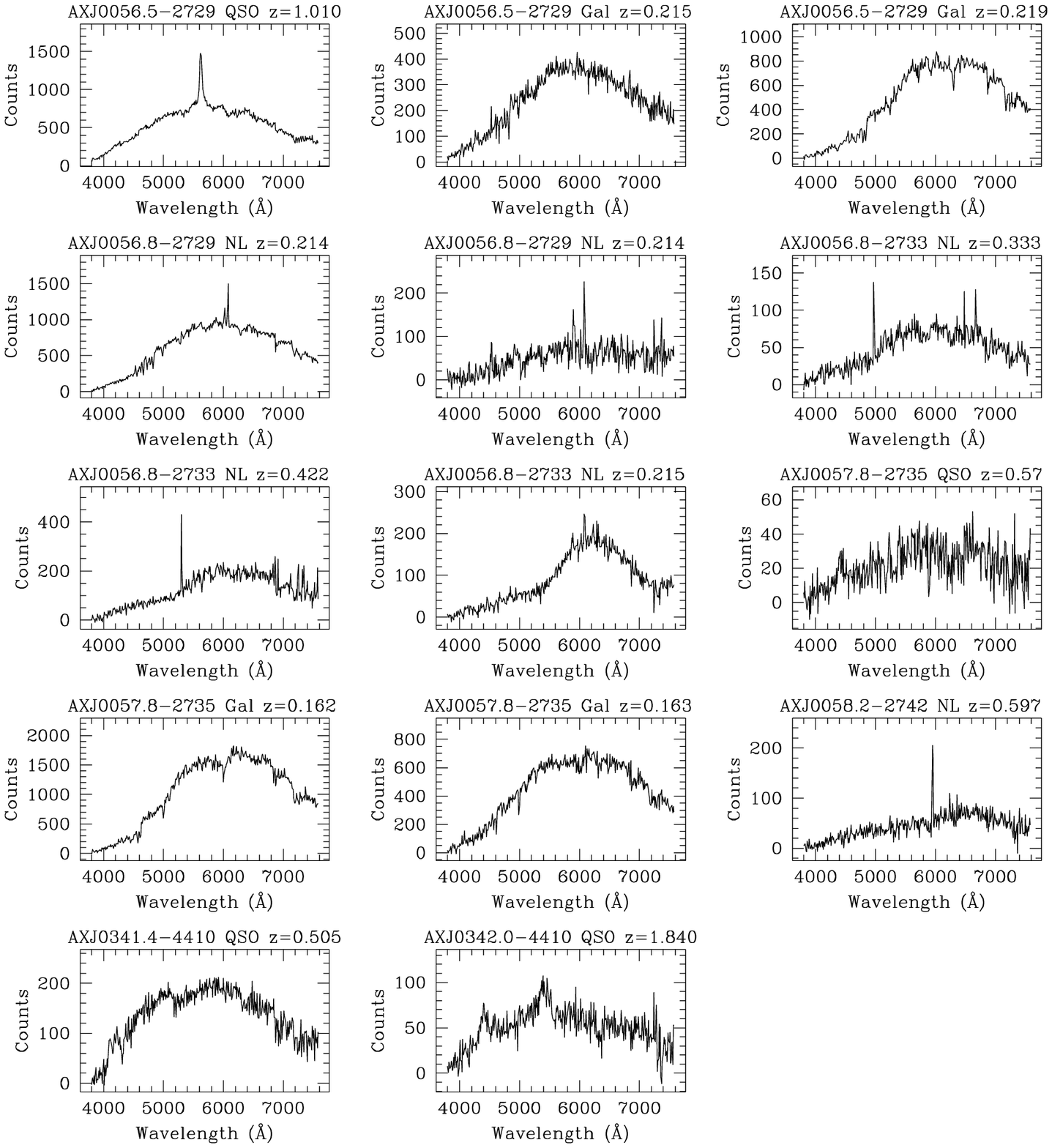,width=6.4in}\break
\noindent\bf Figure 1. \rm LDSS spectra of the optical 
counterparts to the {\it ASCA} sources.}

\subsection{Optical}\tx

In Paper I, we were able to present a preliminary spectroscopic
identification of the optical counterparts to the {\it ASCA} sources,
based on those already identified as counterparts to the corresponding
deep {\it ROSAT} sources in the cross-correlation.  However, the
optical identification of the $4\sigma$ list in the deep {\it ROSAT}
survey was not complete, and this led to significant incompleteness in
the optical identification ($\sim 50\,$per cent) for the {\it ASCA}
source list.

\table{1}{D}{\noindent\bf Table 1.\ \ \rm {\it ASCA} Survey Catalogue}
{\tabskip=1em plus 2em minus .7em 
\halign to\hsize{#\hfil&
\hfil#\hfil&
\hfil#\hfil&
\hfil#&
\hfil#&
\hfil#\hfil&\hfil#&#\hfil\cr
\noalign{\medskip}
\noalign{\hrule}
\noalign{\smallskip}     
&  {\it ASCA} position  &  Optical Position & \multispan2 \hfil Offset \hfil & Count rate \cr
Name &  RA  (2000) Dec &  RA (2000) Dec  & {\it ASCA}&{\it ROSAT}&  2-10keV &\hfil B \hfil&  Spectroscopic ID \cr
     &  h\quad  m\quad  s\qquad  $^{\circ}$\quad  $'$\quad $''$\quad  &  h\quad  m\quad  s\qquad  $^{\circ}$\quad  $'$\quad $''$\quad
& ($''$)  & ($''$)  & ($10^{-3}\,$ct$\,$s$^{-1}$)&(mag)\cr
\noalign{\smallskip}
\noalign{\hrule}
\noalign{\smallskip}     
AX J0056.4-2748            & 00 56 25.6  $-$27 48 48 &  00 56 23.6  $-$27 48 56 &  27 &  6 &5.07$\pm$1.24 & 16.0 &  {\bf QSO z=0.145}\cr
AX J0056.5-2729            & 00 56 31.1  $-$27 29 47 &  00 56 34.0  $-$27 29 52 &  39 & 17 &3.30$\pm$1.14 & 16.9 & {\bf QSO z=1.010}$^{*,1}$\cr
                           &                         &  00 56 34.0  $-$27 29 52 &  39 & 17 &3.30$\pm$1.14 & 16.9 &  Galaxy z=0.215$^{*,1}$\cr
                           &                         &  00 56 32.6  $-$27 29 59 &  23 & 22 &3.30$\pm$1.14 & 19.3 &  Galaxy z=0.219$^{*,1}$ \cr
AX J0056.8-2729$^{\P}$     & 00 56 49.9  $-$27 29 28 &  00 56 50.3  $-$27 29 34 &   8 &    &1.90$\pm$0.88 & 23.0 &  {\bf NL z=0.214}$^{*,2}$ \cr
                           &                         &  00 56 51.7  $-$27 28 57 &  39 &    &1.90$\pm$0.88 & 18.1 &  NL z=0.214$^{*,2}$ \cr
AX J0056.8-2733            & 00 56 51.0  $-$27 33 09 &  00 56 50.0  $-$27 33 21 &  18 &  5 &1.66$\pm$0.60 & 22.7 &  {\bf NL z=0.333}$^{*,3}$\cr
                           &                         &  00 56 52.2  $-$27 33 39 &  34 & 31 &1.66$\pm$0.60 & 21.6 &  NL z=0.422$^*$\cr
                           &                         &  00 56 48.8  $-$27 33 31 &  37 & 16 &1.66$\pm$0.60 & 21.9 &  NL z=0.215$^*$ \cr
AX J0057.0-2741            & 00 56 59.4  $-$27 41 00 &  00 56 56.9  $-$27 40 30 &  45 &  8 &1.06$\pm$0.46 & 21.9 &  {\bf Cluster z=0.561}$^4$\cr
AX J0057.0-2741${\ddag}$   &                         &  00 57 04.4  $-$27 40 22 &  76 & 15 &1.06$\pm$0.46 & 21.6 &  Continuum\cr
AX J0057.3-2731            & 00 57 20.8  $-$27 31 53 &  00 57 24.4  $-$27 32 01 &  49 &  8 &3.07$\pm$0.36 & 19.5 &  {\bf QSO z=1.209}\cr
AX J0057.6-2731$^{\dag}$   & 00 57 38.5  $-$27 31 13 &  00 57 45.6  $-$27 31 17 &  96 & 37 &1.13$\pm$0.44 & 20.1 &  {\bf NL z=0.316}$^5$\cr
AX J0057.8-2735            & 00 57 48.4  $-$27 35 56 &  00 57 46.8  $-$27 35 37 &  28 &  3 &0.87$\pm$0.37 & 22.5 &  {\bf QSO z=0.57}$^{*,6}$\cr
                           &                         &  00 57 48.0  $-$27 35 54 &   6 & 27 &0.87$\pm$0.37 & 18.9 &  Galaxy z=0.162$^{*,6}$\cr
                           &                         &  00 57 49.4  $-$27 36 22 &  29 & 60 &0.87$\pm$0.37 & 18.0 &  Galaxy z=0.163$^{*,6}$\cr
AX J0058.2-2742            & 00 58 11.9  $-$27 42 44 &  00 58 13.5  $-$27 42 11 &  39 & 16 &1.70$\pm$0.64 & 22.2 &  {\bf NL z=0.597}$^*$\cr
AX J0341.1-4412            & 03 41 04.5  $-$44 12 04 &  03 41 04.2  $-$44 12 06 &   4 &  8 &1.19$\pm$0.36 & 21.6 &  {\bf QSO z=1.808}\cr
AX J0341.4-4410            & 03 41 23.0  $-$44 10 47 &  03 41 19.3  $-$44 10 30 &  43 &  7 &1.01$\pm$0.29 & 21.5 &  {\bf QSO z=0.505}$^*$\cr
AX J0341.8-4414$^{\P}$     & 03 41 45.6  $-$44 14 07 &                          &     &    &1.00$\pm$0.31 &      &  {\bf No reliable c'part} \cr
AX J0341.8-4402            & 03 41 47.1  $-$44 02 24 &  03 41 44.3  $-$44 02 49 &  39 & 14 &0.92$\pm$0.31 & 21.7 &  {\bf QSO?}$^7$\cr
AX J0341.8-4353            & 03 41 51.2  $-$43 53 48 &  03 41 52.2  $-$43 53 26 &  24 &  7 &3.00$\pm$0.54 & 12.0 &  {\bf G star}\cr
AX J0342.0-4410$^{\dag}$   & 03 42 01.1  $-$44 10 53 &  03 42 05.6  $-$44 09 43 &  85 &  7 &1.26$\pm$0.28 & 21.1 &  {\bf QSO z=1.840}$^{*,8}$\cr
AX J0342.0-4403            & 03 42 02.4  $-$44 03 51 &  03 42 03.7  $-$44 03 45 &  15 &  3 &0.99$\pm$0.30 & 19.2 &  {\bf QSO z=0.635} \cr
AX J0342.3-4412            & 03 42 19.1  $-$44 12 38 &  03 42 18.4  $-$44 12 50 &  14 &  5 &0.82$\pm$0.31 & 21.8 &  {\bf QSO z=1.091} \cr
AX J0342.3-4412$^{\ddag}$  &                         &  03 42 16.7  $-$44 11 54 &  52 &  2 &0.82$\pm$0.31 & 21.7 & Not observed\cr
AX J0342.5-4409$^{\dag}$   & 03 42 32.4  $-$44 09 35 &  03 42 24.1  $-$44 09 44 &  90 & 11 &0.92$\pm$0.37 & 22.1 &  {\bf Not observed} \cr
                 &                                   &  03 42 24.6  $-$44 09 50 &  85 & 14 &0.92$\pm$0.37 & 20.6 &  Not observed   \cr
AX J0342.6-4404            & 03 42 35.4  $-$44 04 41 &  03 42 38.5  $-$44 04 50 &  34 &  1 &1.39$\pm$0.42 & 20.4 &  {\bf QSO z=0.377}  \cr
AX J1046.1-0020            & 10 46 05.1  $-$00 20 48 &  10 46 05.9  $-$00 20 25 &  26 & 12 &1.54$\pm$0.42 & 20.6 & {\bf QSO z=1.070} \cr
AX J1046.1-0020$^{\ddag}$  &                         &  10 46 05.2  $-$00 22 02 &  74 & 14 &1.54$\pm$0.42 & 20.6 & NL z=0.295 \cr
AX J1046.2-0022            & 10 46 13.4  $-$00 22 16 &  10 46 15.0  $-$00 22 48 &  40 & 13 &1.26$\pm$0.30 & 20.8 & {\bf QSO z=1.952} \cr
AX J1046.4-0021$^{\dag}$   & 10 46 26.6  $-$00 21 09 &  10 46 33.6  $-$00 21 40 & 109 & 20 &1.14$\pm$0.34 & 18.5 & {\bf NL  z=0.130} \cr
AX J1046.7-0021$^{\P}$     & 10 46 39.8  $-$00 21 40 &                          &     &    &0.97$\pm$0.37 &      & {\bf No reliable c'part}\cr 
AX J1046.9-0026$^{\dag}$   & 10 46 54.4  $-$00 26 42 &  10 46 53.4  $-$00 25 44 &  60 &  4 &1.84$\pm$0.60 & 21.3 & {\bf NL  z=0.435} \cr
AX J1047.1-0025$^{\dag}$   & 10 47 09.3  $-$00 25 39 &  10 47 12.3  $-$00 26 18 &  60 & 19 &2.65$\pm$0.30 & 22.1 & {\bf Low S/N}  \cr 
AX J1047.2-0028            & 10 47 12.4  $-$00 28 00 &  10 47 14.8  $-$00 28 29 &  46 & 27 &4.17$\pm$1.27 & 18.2 & {\bf Galaxy z=0.080}  \cr
\noalign{\smallskip}
\noalign{\hrule}
\noalign{\smallskip}
\noalign{\noindent \bf Notes}
\noalign{\noindent Entries in bold refer to the optical identification
adopted in initial sample.} 
\noalign{\noindent $^*$ Observed with LDSS}
\noalign{\noindent $^{\dag}$ Association with identified {\it ROSAT} 
counterpart ignored in the revised sample.}
\noalign{\noindent $^{\ddag}$ Alternative {\it ROSAT} identification
associated with {\it ASCA} source.  Optical counterpart ignored in the
revised sample.}
\noalign{\noindent $^{\P}$ No {\it ROSAT} counterpart within 90arcsec.}
\noalign{\noindent 1.\ A QSO and a $z \sim 0.215$ galaxy cluster are 
both detected as counterparts to the same {\it ASCA} source.}
\noalign{\noindent 2.\ Two galaxies observed at the same redshift.  
They may comprise a possible cluster, but no obvious cluster is seen 
in the direct image. Both galaxies have emission-line ratios 
consistent with Seyfert 2 galaxies (see text).}
\noalign{\noindent 3.\ There are several narrow-emission-line 
galaxies identified within the {\it ASCA} error box.  However, the 
$z=0.333$ object closest to the corresponding {\it ROSAT} centroid is 
most likely to be the correct identification.  Its emission-line 
ratios are consistent with a Seyfert 2 galaxy.}
\noalign{\noindent 4.\ Cluster identification from Couch et al. (1985).}
\noalign{\noindent 5.\ The {\it ROSAT} and {\it ASCA} source positions 
are widely separated.  The $z=0.316$ narrow-emission line galaxy was 
originally identified as a counterpart to the {\it ROSAT} source, but
it exhibits a large separation from both the {\it ROSAT} and {\it
ASCA} centroids. The source is coincident with the outer regions of
one of the spiral arms of the brightest galaxy $B=13$ in the GSGP4
field (GSA002, see Peterson et al.\ 1986.  However, the {\it ASCA}
source is unlikely to be associated with the galaxy ($z=0.019$) given
the implied X-ray-to-optical flux ratio for a galactic source at this
redshift.}
\noalign{\noindent 6.\ The closest object to the {\it ROSAT} centroid 
is a probable $z=0.57$ QSO observed at low signal-to-noise.  The
closest object to {\it ASCA} centroid is a $z=0.162$ galaxy, with an
additional galaxy at the same redshift some 20 arcsec further away
from the {\it ROSAT} position.}
\noalign{\noindent 7.\ Radio source: $S_{\rm 20cm}$=298mJy}
\noalign{\noindent 8.\ Radio source: $S_{\rm 20cm}$=189mJy}   
}}

\table{2}{S}{\noindent\bf Table 2.\ \ \rm Identification
summary for the {\it ASCA} survey}
{\tabskip=1em plus 2em minus .7em 
\halign to\hsize{#\hfil&
\hfil#\hfil&
\hfil#\hfil\cr
\noalign{\medskip}
\noalign{\hrule}
\noalign{\smallskip} 
Classification&\multispan2\hfil Number \hfil\cr
&Initial&Revised\cr
\noalign{\medskip} 
\noalign{\hrule}
\noalign{\smallskip} 
QSO&13&10\cr
Narrow-emission line objects&6&1\cr
Cluster&1&4\cr
Star&1&1\cr
Absorption-line galaxy&1&1\cr
No identification&4&9\cr
\noalign{\medskip}
\noalign{\hrule}
\noalign{\smallskip} 
}}

In order to improve the completeness of the optical identifications
for the {\it ASCA} source list, we carried out spectroscopic
observations of optical counterparts to {\it ASCA}/{\it ROSAT} sources
using the Low Dispersion Survey Spectrograph (LDSS) at the
Anglo-Australian Telescope (AAT) on the nights of 1996 November
8--10.  We operated LDSS using a mask comprising of a single
`long-slit' (1.7$\,$arcsec wide) centred at 5500\AA, with the high
dispersion grism giving an overall
instrumentation resolution of 13\AA\ over the wavelength range
$3800<\lambda<7800$.  Throughout the run the conditions were less than
ideal, with seeing ranging from 1.5 arcsec to over 3 arcsec; but all
identifications reported here were obtained in less than 2.5 arcsec
seeing.  Little or no photometric time was obtained, and observations
were only possible on the equivalent of 1.5 nights out of the 3
allocated.  With LDSS, the slit mask and grism can be removed to give
direct imaging, and we also obtained direct (unfiltered) images for
most of the optical counterparts we observed.  In most cases, the LDSS
slit was oriented so that 2 or more optical counterparts could be
positioned along the slit.

Optical counterparts for observation with LDSS were selected from APM
scans of the UK Schmidt telescope (UKST) J survey plates and AAT J
plates in each of the 2 {\it ASCA} areas (GSGP4 and QSF3) observable
during the observing run.  Candidates were simply selected on the
basis on their proximity from the centroid of the {\it ROSAT} source
which was the counterpart to the {\it ASCA} source.  Orientation of
the LDSS slit-mask allowed us to observe two or more optical
counterparts for each integration.  For each X-ray source, we
continued to obtain spectra for optical counterparts until a plausible
identification (e.g. QSO, cluster, emission-line object) was obtained.

Based on these observations, the revised identification list for the
{\it ASCA} sources is presented in Table 1.  {\it ASCA} sources with
spectroscopic identifications from the LDSS observations are indicated
with an asterisk.  LDSS spectra for these objects are presented in
Fig.\ 1. Since the slit was not positioned at the parallactic angle,
and the seeing disk was typically much wider than the slit width, we
chose not to flux the LDSS spectra, and they are presented here simply
with counts on the ordinate.  AUTOFIB spectra for the remainder of 
the sources
(identified as part of the {\it ROSAT} deep survey) are published in
Almaini (1996), or Shanks et al. (in preparation).  Notes for some of
the identifications are given at the foot of Table 1.

Following the LDSS observations, we now have an optical identification
for up to 22 of the 26 {\it ASCA} sources, a completeness rate of 85
per cent.  The breakdown of these initial classifications is
summarised in Table 2.  Of these 22 sources, there are 13 QSOs, 6
objects with narrow emission lines, 1 cluster, 1 star and 1
absorption-line galaxy. All but 2 of these sources, AXJ0057.6$-$2731
(a narrow-emission line galaxy) and AXJ1047.2$-$0028 (an absorption
line galaxy) were found within 20$\,$arcsec of the {\it ROSAT} X-ray
position.  At our optical detection limit on the UKST photographic
plates ($B\sim23$), the space density of QSOs is $\sim
200\,$deg$^{-2}$ (Schade et al.\ 1996).  Thus we would expect 0.02
QSOs within 20 arcsec of each X-ray source position by chance,
corresponding to only 0.5 QSOs over the entire survey.
 
It is difficult to make a definitive identification of the nature
of the six narrow-emission-line galaxies identified in this survey.
In the four cases where it is possible to make a reliable 
measurement of the ${\rm [OIII]\lambda5007/H\beta}$ ratio, all but one
(the counterpart to AXJ0057.6--2731) have emission-line ratios 
${\rm [OIII]\lambda5007/H\beta > 3}$.  Based on the standard 
Baldwin, Phillips and Terlevich (1981) line-ratio
diagnostic, this observation in itself is insufficient to classify
these objects unambiguously as starburst galaxies or Seyfert 2
galaxies.  Other line ratios, most notably [NII]/H$\alpha$, are also
required.  However, in the detailed study of narrow-emission line
galaxies identified as counterparts to {\it ROSAT} X-ray sources by
Boyle et al.\ (1995), the vast majority (7/8) of the narrow-emission
line galaxies with ${\rm [OIII]\lambda 5007/H\beta > 3}$, also have
[NII]/H$\alpha$ emission line ratios which identify them
as Seyfert 2 galaxies rather than starburst galaxies 
(see Fig 3.\ in Boyle et al.\ 
1995).  The 2-10$\,$keV X-ray luminosities of the 
narrow-emission-line galaxies identified in this survey 
($>10^{42.5}\,$\ergs) are also inconsistent with starburst 
galaxies.  We tentatively conclude, therefore, that 
the majority of narrow-emission-line objects identified in this 
survey are more likely to be Seyfert-2 galaxies (Type 2 AGN) than
starburst galaxies.  However, further observations will be required
to determine their nature unequivocally.

Tresse et al.\ (1996) estimate that approximately 17 per cent of field
galaxies with $I<22.5$ have emission line ratios consistent with
active galaxies (Seyfert 2 or Liners).  Based on the galaxy
number-magnitude counts of Metcalfe et al.\ (1991), this corresponds
to a surface density of 800 active galaxies per square degree
at $B<23$, implying that a total of $\sim 2$ such objects would be 
found by chance amongst the optical counterparts found within 
20$\,$arcsec of
the {\it ROSAT} source positions associated with {\it ASCA}
detections.  In total, seven narrow-emission-line galaxies were identified at
these separations, including a narrow-emission-line galaxy found only
8$\,$arcsec from an {\it ASCA} source position with no {\it ROSAT}
counterpart.  Of these seven sources, five have been tentatively
identified as the counterpart to an {\it ASCA} source, the remaining
two are the more distant optical counterparts to {\it ASCA} sources
with an alternative identification (AXJ0056.8$-$2733 and
AXJ1046.1$-$0020).  This confusion rate is therefore broadly
consistent with that predicted from the observations of 
Tresse et al.\ (1996).

There is one narrow-emission-line galaxy (AXJ0057.6-- 2731) identified
as a counterpart to an {\it ASCA} source which was not found within
20$\,$arcsec of the corresponding {\it ROSAT} source position. It is
probable that this identification (a late-type galaxy with weak [OII]
and [OIII] emission) is not the correct counterpart to the {\it ASCA}
source.  This galaxy is 34$\,$arcsec and 95$\,$arcsec distant from the
{\it ROSAT} and {\it ASCA} source positions respectively and the
separation between {\it ROSAT} and {\it ASCA} positions is also large
($>60\,$arcsec).  It is therefore one of the positional coincidences
between the {\it ASCA} and {\it ROSAT} catalogues most likely to have
been chance and its identification was one of those removed from the
revised catalogue (see below).

There may be other optical identifications that are also 
ambiguous.  We have included in the list of QSOs at least 2 sources
whose classification is uncertain.  One {\it ASCA} source,
AXJ0341.8--4402 has been associated with a faint ($B=21.7$) blue stellar
object 14 arcsec distant from the {\it ROSAT} position and 39 arcsec
from the {\it ASCA} centroid.  We were frustrated in our attempts to
obtain a spectrum for this object by the weather conditions.  From
observations made with the Australia Telescope (Boyle et al.\ 1993)
this object is also a relatively strong radio-source ($S_{\rm
20cm}=298\,\mu$Jy).  Thus the most likely identification for this
source is a QSO, although it will require further confirmation.  The
closest optical counterpart to the {\it ROSAT} source coincident with
AXJ0057.8--2735 is identified as a $z=0.57$ QSO based on a low
signal-to-noise spectrum containing one emission line.  It is possible
that the correct optical counterpart to the {\it ASCA} source is one
of the pair of $z=0.162$ galaxies, one of which is only 6 arcsec from
the {\it ASCA} centroid.  However, both these galaxies are
$>25\,$arcsec from the {\it ROSAT} position and both are 
unlikely to be the correct identification if the {\it ROSAT} and 
{\it ASCA} detections correspond to the same source.  Finally, the QSO identified as the
counterpart to AXJ0056.5--2729 is a rather remarkable object.
Originally identified as a cluster in Paper I, the LDSS observations
revealed a $z=1.01$ QSO lying behind the cluster at $z=0.217$.  It is
possible that both the cluster and the QSO contribute to the X-ray
emission.

Some of the {\it ASCA} sources initially associated with
narrow-emission-line galaxies may also have been incorrectly
identified. For example, although the {\it ASCA} source
AXJ0056.8--2729 has been identified as a $z=0.214$
narrow-emission-line galaxy located only 8 arcsec from the {\it ASCA}
centroid, a narrow-emission-line galaxy with an identical redshift was
also observed 39 arcsec from the {\it ASCA} source.  The X-ray
emission may therefore originate from a cluster or group which
comprises these two galaxies (possibly the outer regions of the
$z=0.215$ cluster associated with AXJ0056.5--2729 some 3--4 arcmins
distant).  However, no strong concentration of galaxies in visible on
the LDSS direct image of this field.

As described above, there is a probability that at least some of the
{\it ROSAT}/{\it ASCA} correlations have come about through chance
coincidences.  The {\it ROSAT}/{\it ASCA} coincidences listed in table
1 are based on separations up to 90 arcsecs.  At this separation,
there are 26 observed coincidences between {\it ROSAT} and {\it ASCA}
sources (3 {\it ASCA} sources have two {\it ROSAT} counterparts at
this separation and 3 {\it ASCA} sources have no {\it ROSAT}
counterpart), of which $\sim 9$ will be random coincidence (see Paper
I).

In order to account for this in the analysis below, we can remove from
the identification list the counterparts to the 9 {\it ROSAT} sources
sources with the largest {\it ASCA}-{\it ROSAT} separation.  This
results in the rejection of the additional {\it ROSAT} counterparts to
{\it ASCA} sources AXJ0057.0--2741, AXJ0342.3--4412 and
AXJ1046.1--0020.  It also removes the spectroscopically identified
counterparts to AXJ0057.6-2731 (a $z=0.316$ narrow-emission-line
galaxy), AXJ0342.0--4410 ($z=1.840$ QSO), AXJ0342.5--4409 (not
observed) AXJ1046.4--0021 ($z=0.130$ narrow-emission-line galaxy),
AXJ1046.9--0026 ($z=0.43$ narrow-emission-line galaxy) and
AXJ1047.1--0025 (an object with a low signal-to-noise ratio spectrum).
This leaves no {\it ASCA}-{\it ROSAT} coincidences with
separations larger than 1 arcmin.  If we further assume that the 
X-ray sources AXJ0057.8--2735 and AXJ0056.5-2729 classified as 
clusters rather than QSOs and that
AXJ0056.8--2729 should be classified as a cluster rather than a
narrow-emission-line galaxy, then the revised composition of the {\it
ASCA} sources is 10 QSOs, 1 narrow-emission-line galaxy, 4 clusters, 1
star and 1 galaxy.  In this case there are no reliable identifications
for a further 9 objects; a completeness rate of 65 per cent.

The composition of the {\it ASCA} source list for the initial and
revised identification lists is summarised in table 2.  It is likely
that the true composition of the {\it ASCA} source list will lie
somewhere between these two extremes.  Although we were `optimistic'
in assigning identifications to some chance {\it ROSAT}/{\it ASCA}
positional coincidences in our initial sample, we were
`pessimistic' (particularly with regard to emission-line objects) in
our identification rates in the revised list.  In the analysis that
follows, we will use both the initial and revised lists to determine
the likely systematic errors in any estimation of the properties of
the AGN population that forms the bulk of the {\it ASCA} sample, 
whichever list is chosen.

\section{Analysis}\tx

\figure{2}{S}{0mm}{
\psfig{figure=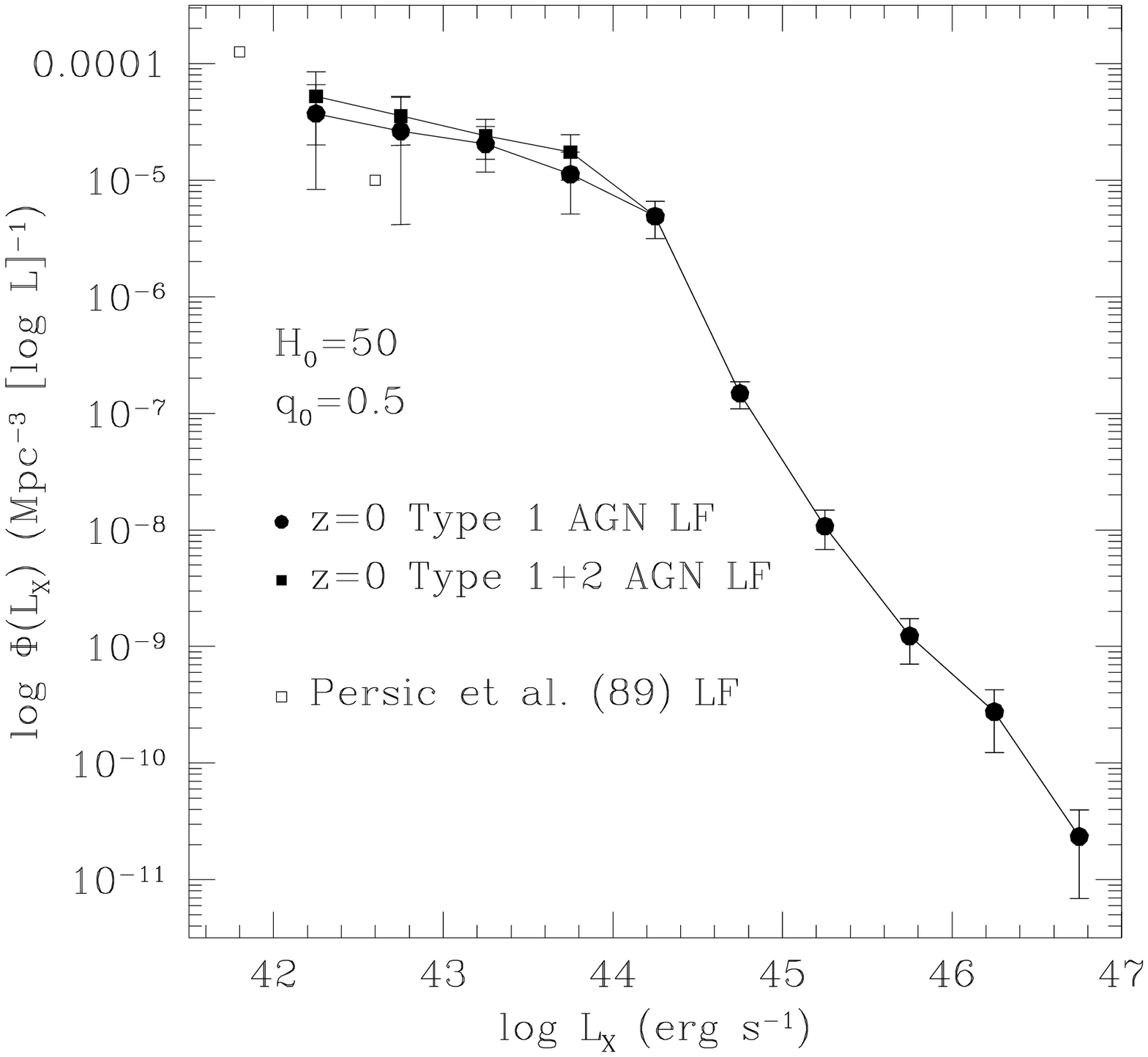,width=3.2in}\break
\noindent\bf Figure 2. \rm $1/V_a$ estimate for the 2-10$\,$keV
luminosity function for Type 1 and Type 1+2 AGN.}

For the purposes of the following analysis, we have combined the
emission-line objects identified in our {\it ASCA} source list (QSOs
and narrow-emission-line galaxies) with the bright sample of AGN
identified by Grossan (1992) from the Large Area Sky 
Survey/Modulation Collimator (LASS/MC) catalogue (Remillard et al.\
1986) obtained from {\it HEAO 1} mission (Wood et al.\ 1984).  The LASS/MC AGN (LMA)
sample comprises 96 predominantly low redshift ($z<0.2$) AGN (85
Seyfert 1 and 11 Seyfert 2 galaxies), selected at bright flux limits
($S_{\rm 2-10} > 2 \times 10^{-11}\,$\ergcms) and is 90 per cent
complete at galactic latitudes $b>20^{\circ}$.  This sample has also
previously been used by Ceballos and Barcons (1996) to derive the
$z<0.2$ AGN luminosity function in the 2-10$\,$keV band.  Combining
these two samples gives us a wide range in luminosity and redshift
over which to determine the evolution of the AGN luminosity 
function (LF).

We carried out the analysis for two AGN samples: Type 1 AGN and Type 1
+ Type 2 AGN.  The Type 1 AGN sample comprises the QSOs identified in
the {\it ASCA} survey and the Seyfert 1 galaxies in the Grossan (1992)
sample.  The Type 1 + Type 2 AGN sample also includes the
narrow-emission-line galaxies identified in the {\it ASCA} sample
(tentatively identified as Seyfert 2-like objects) and the bona-fide
Seyfert 2 galaxies in Grossan (1992). Unfortunately, there are too few
Type 2 AGN to carry out a meaningful analysis of the LF for this
sample alone.  For ease of reference the samples will be referred to
as the Type 1 AGN and Type 1 + Type 2 AGN in the analysis below.
However, we note there is still uncertainty over the identification of
the six narrow-emission-line galaxies in the {\it ASCA} sample as
Seyfert 2 galaxies.

\table{3}{S}{\noindent\bf Table 3.\ \ \rm $\langle{V_e\over V_a}\rangle$ values derived for
AGN in the {\it ASCA} and LMA samples.}
{\tabskip=1em plus 2em minus .7em 
\halign to\hsize{#\hfil&\hfil#&
\hfil#\hfil&\hfil#&
\hfil#\hfil\cr
\noalign{\medskip} 
\noalign{\hrule}
\noalign{\smallskip} 
AGN&\multispan 2\hfil Initial\hfil&\multispan2\hfil Revised\hfil\cr
&N&$\langle{V_e\over V_a}\rangle$&N&$\langle{V_e\over V_a}\rangle$\cr
\noalign{\medskip} 
\noalign{\hrule}
\noalign{\smallskip} 
Type 1 + Type 2&115&$0.63\pm0.03$&108&$0.62\pm0.03$\cr
Type 1&98&$0.64\pm0.03$&95&$0.62\pm0.03$\cr
\noalign{\medskip}
\noalign{\hrule}
\noalign{\medskip}
}}

\table{4}{D}{\noindent\bf Table 4.\rm \ \ The 2-10$\,$keV AGN LF parameters}
{\tabskip=1em plus 2em minus .7em 
\halign to\hsize{#\hfil&
\hfil#\hfil&
\hfil#&
\hfil#&
\hfil#&
\hfil#&
\hfil#&
\hfil#\hfil&
#\hfil\cr
\noalign{\medskip}
\noalign{\hrule}
\noalign{\smallskip}     
Sample&AGN&N&$\gamma_1$&$\gamma_2$&$\log L^*$&k&$\Phi^*$&$P_{KS}$\cr
&&&&&&&Mpc$^{-3}$[$10^{44}\,$erg$\,$s$^{-1}$]$^{-1}$\cr
\noalign{\medskip}
\noalign{\hrule}
\noalign{\smallskip}     
LMA + ASCA initial&Type 1 + Type 2&115&1.91&2.98&44.13&2.04&$9.52\times10^{-7}$&0.20\cr
LMA + ASCA revised&Type 1 + Type 2&108&1.89&2.97&44.14&2.03&$9.19\times10^{-7}$&0.22\cr
LMA + ASCA initial&Type 1&98&1.73&2.96&44.15&2.01&$8.24\times10^{-7}$&0.27\cr
LMA + ASCA revised&Type 1&95&1.73&2.96&44.16&2.00&$8.21\times10^{-7}$&0.25\cr
\noalign{\smallskip}
\multispan2 68\% confidence regions\hfil &&$^{+0.14}_{-0.20}$&$^{+0.06}_{-0.08}$&$^{+0.10}_{-0.10}$&$^{+0.16}_{-0.22}$\cr
\noalign{\smallskip}
\multispan2 95\% confidence regions\hfil &&$^{+0.20}_{-0.70}$&$^{+0.18}_{-0.25}$&$^{+0.34}_{-0.61}$&$^{+0.34}_{-0.46}$\cr
\noalign{\medskip}
\noalign{\hrule}
\noalign{\smallskip}     
}}

We first carried out a standard $\langle{V_e\over V_a}\rangle$
analysis (Avni and Bahcall 1980) of the combined LMA plus {\it ASCA}
samples.  For consistency with previous analyses, the `canonical'
value of $\alpha_X=0.7$ (Mushotzky 1982) was adopted for the AGN
energy spectral index in the 2-10$\,$keV band.  This is marginally
softer than, but not inconsistent with, the mean AGN 2-10$\,$keV
spectral index measured in our sample $\alpha_X\sim0.5\pm0.2$ (see
Georgantopoulos et al.\ 1997, in preparation, hereinafter Paper III).
The adoption of $\alpha_X=0.7$ makes little or no difference to the
computed $\langle{V_e\over V_a}\rangle$; $\alpha_X=0$ and
$\alpha_X=1.0$ change it by no more than $\pm 0.01$.  Similarly, any
change in $\alpha_X$ results in the identical change in the derived
exponent of power-law luminosity evolution ($k$, see below) and
estimates of the AGN contribution to the X-ray background are
independent of $\alpha_X$.

The $\langle{V_e\over V_a}\rangle$ results are reported in Table 3,
where the error on $\langle{V_e\over V_a}\rangle$ is estimated from
$1/\sqrt{12N}$.  All entries in Table 3 are at least $4\sigma$ higher
than the `no-evolution' $\langle{V_e\over V_a}\rangle=0.5$ value all
samples.  There is no significant difference between any of the
$\langle{V_e\over V_a}\rangle$ values listed in this table, whether
they were derived for the Type 1 or Type 1 + Type 2 AGN samples or
with the inclusion of the initial or revised {\it ASCA} samples.
These values for $\langle{V_e\over V_a}\rangle$ are consistent with
that already derived for the AGN in the LMA sample by Grossan (1992);
$\langle{V_e\over V_a}\rangle=0.63$.  Grossan (1992) was concerned
that this high value of $\langle{V_e\over V_a}\rangle$ might be due to
systematic errors in the LMA sample, (as opposed to evolution), but
was unable to identify a potential source of error which could give
rise to this high value of $\langle{V_e\over V_a}\rangle$.  However,
we have also obtained similar values of $\langle{V_e\over V_a}\rangle$
(albeit with larger errors) when we combined either the initial or
revised {\it ASCA} sample with the more limited sample of 26 AGN in
Piccinotti et al. (1982); $0.61\pm 0.05 < \langle{V_e\over V_a}\rangle
< 0.65\pm 0.04$ or when we analysed the {\it ASCA} sample on its own
$0.59\pm 0.09 < \langle{V_e\over V_a}\rangle < 0.65\pm 0.08$.  We
therefore conclude that there is significant evidence for AGN
evolution in the 2-10$\,$keV band.

We then carried out a maximum likelihood analysis to derive the
best-fit parameters of a number of models used to describe the
2-10$\,$keV luminosity function and its cosmological evolution with
redshift.

We chose a two-power-law representation for the QSO LF, identical
to the model used to fit successfully the 0.5-2$\,$keV QSO LF
by Boyle et al.\ (1994):

$$\Phi(L_X) = K_1 L_{X}^{-\gamma _1}\ \ \ \ \ L_X < L_X^*(z=0)$$
$$\Phi(L_X) = K_2 L_{X}^{-\gamma _2}\ \ \ \ \ L_X > L_X^*(z=0)$$ 

\noindent
This model has already been shown to provide a good fit to the
2-10$\,$keV AGN LF by Ceballos and Barcons (1996). Moreover, Grossan
(1992) has already shown that a single-power-law is not an acceptable
fit to this LF.

\noindent
A standard power-law luminosity evolution model was used to 
parameterise the cosmological evolution of this luminosity
function:

$$L_X^*(z) = L_X^*(0)(1+z)^{k}$$

In order to derive the `best-fit' values and errors for the 4 free 
parameters in this model, $\gamma_1$, $\gamma_2$, $k$ and $L^*$,
we used
our standard likelihood minimisation procedure (Boyle et al.\ 1988).
$K_1$ [$=K_2 (L_{X}^*/10^{44}{\rm erg\,s^{-1}})
^{(\gamma _1 - \gamma _2)}$] is obtained from the normalisation of 
the luminosity function to the total number of objects identified in 
the combined sample.

As before, the acceptability of the model was derived from the 2D
Kolmogorov-Smirnov statistic applied to the joint distribution in
luminosity and redshift.

The model was tested against the combined LMA plus initial and revised
{\it ASCA} samples, and the model was run for both the Type 1 and Type
1 + Type 2 AGN samples.  The results are reported in Table 4.  The
rate of the cosmological evolution derived from these fits was also
used to generate the binned $1/V_a$ estimate of the de-evolved LF for
both Type 1 and Type 1 + Type 2 AGN in the initial sample which is
plotted in Fig 2.  The error bars on this figure are based on Poisson
statistics.  The binned estimates of the 2-10$\,$keV LF from Persic et
al.\ (1989) are also plotted on this figure.  The Persic et al.\
(1989) LF is based on {\it HEAO 1} A-2 observations of the complete
sample of Seyferts in the CfA optically-selected sample and provides
an independent estimate of the 2-10$\,$keV LF at low luminosities.  We
have plotted the lower estimates of the LF from Persic et al.\ (1989).
The Persic et al.\ (1989) LF exhibits good agreement with the faint
end of the LF derived in this analysis.

We can see from Table 4 that a two-power-law LF with power-law
luminosity evolution produces a good fit to the data.  There is no
significant difference between the model fits to the initial and
revised {\it ASCA} data-sets.  The main difference between the model
fits to the Type 1 and Type 1 + Type 2 AGN samples is in the slope of
the faint end of the 2-10$\,$keV LF.  As borne out by Fig.\ 2, the
addition of Type 2 AGN into the fit marginally steepens the faint end
of the LF.

The values for parameters of the $z=0$ LF obtained here are different
at the 95 per cent confidence level to those derived by Ceballos and
Barcons (1996); $\gamma_1=2.11$, $\gamma_2=3.27$, and $\log
L^*=44.51$.  Ceballos and Barcons (1996) computed the LF from the
$z<0.2$ AGN in the LMA sample using the evolution derived by Maccacaro
et al.\ (1991) for the EMSS sample in the 0.3-3.5$\,$keV band to
`de-evolve' the LF back to $z=0$.  The EMSS evolution is stronger, $L
\propto (1+z)^{2.6}$, than that derived here, which may, in part,
account for some of the discrepancy seen.

Although the 95 per cent confidence regions on the LF parameter values
($\gamma_1$, $\gamma_2$, $L^*$) quoted in Table 4 are relatively
large, they are highly correlated.  A decrease in the value of
$\gamma_1$ leads to decrease in the derived value of $L^*$ and
corresponding increase in $\gamma_2$. The error on $k$ is largely
uncorrelated with any of the LF parameters, and remains the major
uncertainty in the determination of the AGN contribution to the
2-10$\,$keV X-ray background (see below).

\section{Discussion}\tx

\figure{3}{S}{0mm}{
\psfig{figure=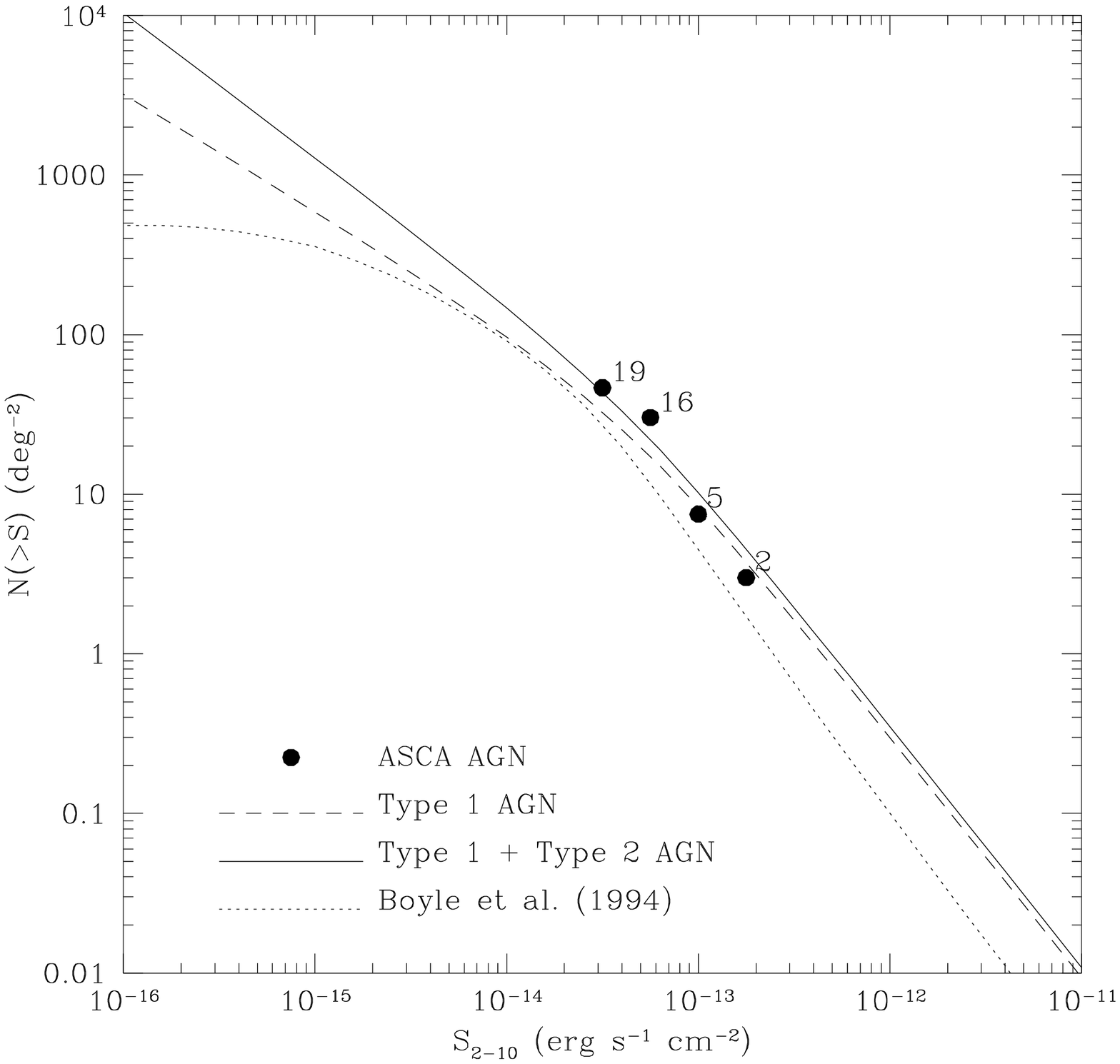,width=3.2in}\break
\noindent\bf Figure 3. \rm The predicted $\log N - \log S$
relations for Type 1 AGN and Type 1 + Type 2 AGN based on the
evolutionary models described in the text.  The surface density of
QSOs and narrow-emission-line galaxies identified in the initial {\it
ASCA} sample are indicated by the filled dots. The predicted
2-10$\,$keV QSO $\log N - \log S$ based on model H in Boyle et al.\
(1994) and extrapolated with $\alpha_X=1$ is shown as a dotted line. }

\figure{4}{S}{-35mm}{
\psfig{figure=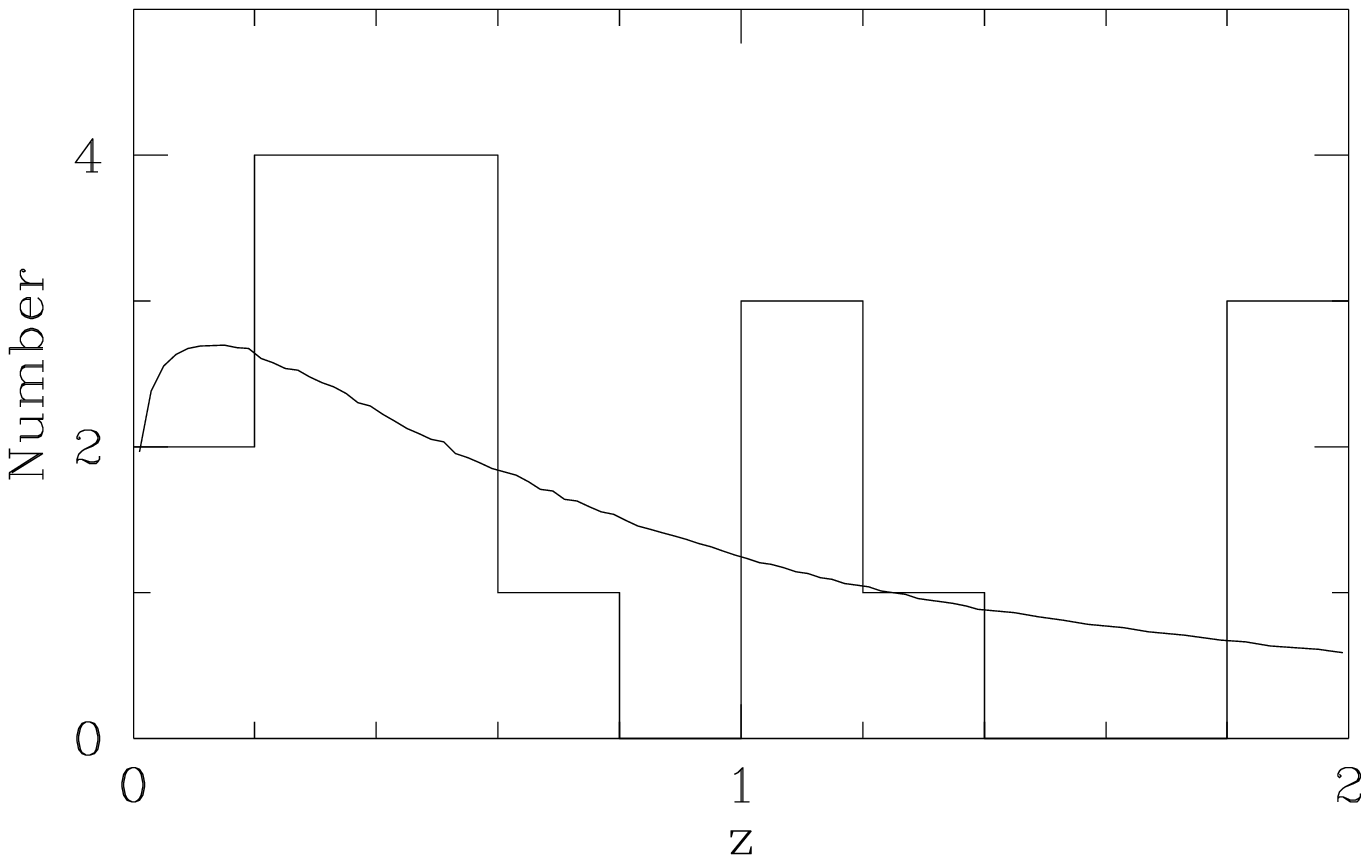,width=3.2in}\break
\noindent\bf Figure 4. \rm The observed and predicted
n(z) relation for Type 1 + Type 2 AGN in the initial {\it ASCA}
sample.}

\figure{5}{S}{-15mm}{
\psfig{figure=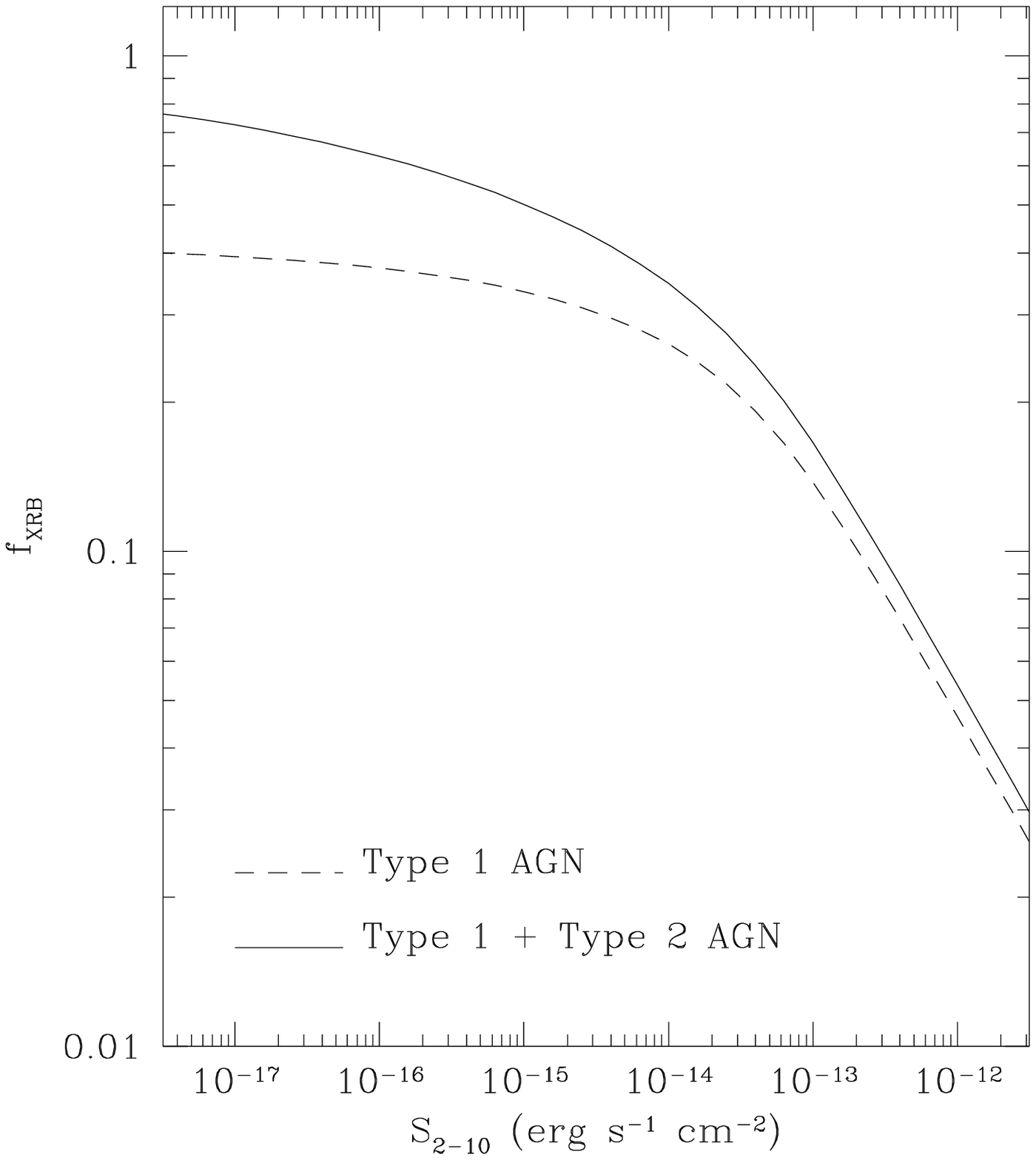,width=3.2in}\break
\noindent\bf Figure 5. \rm The predicted contribution of Type
1 + Type 2 AGN to the 2-10$\,$keV X-ray background, $f_{\rm XRB}$.}

\table{5}{D}{\noindent\bf Table 5.\rm \ \ Predicted AGN contribution,
$f_{\rm XRB}$, to the 2-10$\,$keV X-ray background} 
{\tabskip=1em plus 2em minus .7em
\halign to\hsize{\hfil#\hfil&\hfil#\hfil&\hfil#\hfil&\hfil#\hfil\cr
\noalign{\medskip}
\noalign{\hrule}
\noalign{\smallskip}     
&&\multispan2\hfil $f_{\rm XRB}$ (per cent)\hfil\cr
Limits of integration&LF model&\hfil Type 1\hfil&
\hfil Type 1 + 2\hfil\cr
\noalign{\medskip}
\noalign{\hrule}
\noalign{\smallskip}     
$L>10^{42}\,$erg$\,$s$^{-1}$,\ \ $z<2$&Best-fit&\ \ 27&\ \ 39\cr
$L>10^{42}\,$erg$\,$s$^{-1}$,\ \ $z<4$&Best-fit&\ \ 33&\ \ 48\cr
$L>10^{41}\,$erg$\,$s$^{-1}$,\ \ $z<4$&Best-fit&\ \ 37&\ \ 61\cr
$L>10^{40}\,$erg$\,$s$^{-1}$,\ \ $z<4$&Best-fit&\ \ 39&\ \ 72\cr
$L>10^{39}\,$erg$\,$s$^{-1}$,\ \ $z<4$&Best-fit&\ \ 41&\ \ 80\cr
$L>10^{39}\,$erg$\,$s$^{-1}$,\ \ $z<4$&$\gamma_1+1\sigma$&\ \ 
59&$>100$\cr
$L>10^{39}\,$erg$\,$s$^{-1}$,\ \ $z<4$&$k+1\sigma$&\ \ 46&\ \ 93\cr
\noalign{\medskip}
\noalign{\hrule}
\noalign{\smallskip}     
}}

The predicted 2-10$\,$keV AGN number-flux relation, $\log N - \log S$,
for the two-power-law LF model is plotted in Fig.\ 3 for both Type 1
and Type 1 + Type 2 AGN.  These predicted $\log N - \log S$ relations
were based on an extrapolation the LFs to 
$L=10^{39}\,$erg$\,$s$^{-1}$ and $z=4$ (see below).  
We also show the observed QSO +
narrow-emission-line galaxy $\log N - \log S$ from the {\it ASCA}
survey for the initial sample (corrected for incompleteness).  No
error bars have been included on these points, because they are not
independent.  However the numbers next to each point represent the
number of objects comprising each estimate of the surface density.
The models accurately predict the correct observed numbers in the {\it
ASCA} survey, although the {\it ASCA} survey was of course used, in
part, to determine the parameters of the LF.  The observed and
predicted number-redshift relation, n(z), for Type 1 + Type 2 AGN in
the {\it ASCA} survey are also plotted in Fig.\ 4.  Although the
numbers are small, the model fit is not inconsistent with the data.

We have also used the LF parameters to estimate the contribution of
Type 1 and Type 1 + Type 2 AGN to the total 2-10$\,$keV XRB.  We used
the 2-10$\,$keV XRB, $I=19.5\times10^{-12}\,$\ergcms$\,$deg $^{-2}$\
observed by Chen, Fabian \& Gendreau (1996).  Since the faint end of
the LF is steep, the total contribution of AGN to the XRB is highly
dependant on the faintest X-ray luminosities to which the LF is
extrapolated. To a lesser extent, the total contribution is also
dependent on the redshift to which the evolution of the LF
extrapolated. In Table 5 we list the predicted contributions from AGN
based on the
extrapolations of the Type 1 and Type 1 + Type 2 LFs derived above.  
If we integrate the LF down
to $10^{42}\,$erg$\,$s$^{-1}$ and to $z<2$, the faintest
luminosities and the highest redshifts sampled in the LASS/MC and {\it
ASCA} samples, we obtain a total contribution to the $2-10\,$keV XRB
of 27 per cent for Type 1 AGN and 39 per cent for Type 1 + Type 2 AGN.
In contrast, by extending the integration of the LF to
$10^{39}\,$erg$\,$s$^{-1}$ (the upper limits on the faintest AGN
luminosities in the Persic et al.\  sample), and increasing the
redshift limit to $z=4$ (with no evolution of the LF at $z>2$, see
Boyle et al.\ 1994) we derive total contributions of 41 per cent for
Type 1 AGN and 80 per cent for Type 1 + Type 2 AGN.  When the LF is
extrapolated down to these low luminosities, the error on the slope of
the faint end of the LF dominates the statistical uncertainty in the
estaimate AGN contribution to the XRB.  Increasing the slope of the
faint end of the LF by its $1\sigma$ error yields a total contribution
of almost 60 per cent for Type 1 AGN and over 100 per cent for Type 1
+ Type 2 AGN, saturation occuring at $S_{\rm 2-10} \sim 2.5 \times
10^{-17}\,$\ergcms .  The only other parameter whose statistical
uncertainty significantly affects the AGN contribution to the XRB is
the evolution parameter.  The $1\sigma$ upper limit for $k$ yields
total contributions of 46 per cent and 93 per cent for Type 1 and Type
1 + Type 2 AGN respectively. 

Although upper limits for AGN in  Persic et al.\ (1989)
suggest AGN 2-10$\,$keV luminosities as low as $10^{39}\,$erg$\,$s$^{-1}$, we
note that the local 2-10$\,$keV luminosity density obtained by
integrating the Type 1 + Type 2 $z=0$ LF down to
$10^{42}\,$erg$\,$s$^{-1}$ is
$4.87\times10^{38}\,$erg$\,$s$^{-1}$Mpc$^{-3}$.  This is close to the
95 per cent confidence upper limit of
$4.77\times10^{38}\,$erg$\,$s$^{-1}$Mpc$^{-3}$ obtained for
the local Type 1 +
Type 2 AGN 2-10$\,$keV luminosity density by Barcons et al.\
(1995) from consideration of the {\it HEAO 1} A-2 all-sky maps and the
{\it IRAS} 12 micron source catalogue. Decreasing the minimum
luminosity to $10^{39}\,$erg$\,$s$^{-1}$ in the calculation of the
local 2-10$\,$keV luminosity density yields a value
$8.10\times10^{38}\,$erg$\,$s$^{-1}$Mpc$^{-3}$ which is high compared
to the Barcons et al.\ (1995) upper limit.  

If the AGN LF does cut-off at $L<10^{42}\,$erg$\,$s$^{-1}$,
then we are forced to conclude that there is a significant
residual contribution to the 2-10$\,$keV XRB from sources
which we have yet to identify.  These could be heavily
absorbed AGN (e.g. Madau et al.\ 1994, Comastri et al.\ 1995)
or precursor AGN at high redshift (Boldt and Leiter 1993).

In constrast, if the AGN LF can be extended to significantly fainter
luminosities then AGN could supply most of the 2-10$\,$keV background.
In this case, there is still the outstanding issue of matching the
observed AGN spectrum to that of the X-ray background.  Although the
mean 2-10$\,$keV spectral index for AGN in this analysis
($\alpha_X=0.5\pm0.25$) is consistent with the observed value for the
X-ray background at these energies, it is also equally consistent with
the `canonical' value for low redshift range AGN $\alpha_X=0.7$
(Mushotsky 1982).  The AGN in this analysis could, in principle,
readily exhibit a harder spectrum than $\alpha_X=0.7$ if the
instrinsic AGN spectral index hardened with energy. At the mean
redshift of the sample ($z \sim 0.7$), the observed {\it ASCA} energy
range corresponds to 3.4-17$\,$keV, significantly harder than the
energy range sampled at $z=0$.  Evidence for the hardening of the
X-ray spectral index with energy, possibly due to Compton reflection,
has been found previously (Pounds et al.\ 1990, Nandra \& Pounds,
1994) and has also been used to explain the observed spectrum of the
XRB (Tucker \& Schwarz 1986, Boyle 1996).  However, as yet there is no
strong evidence for any hardening of the X-ray spectral slope with
energy in this sample and further data will be required before a more
accurate picture of the spectral characteristics of these AGN can be
built up.

In Fig.\ 3 we also show the predicted QSO $\log N - \log S$ based on
the fit to the 0.3-3.5$\,$keV XLF made by Boyle et al. (1994, model H)
based on the {\it ROSAT} (0.5-2$\,$keV) and Einstein (0.3-3.5$\,$keV)
surveys and extrapolated to the 2-10$\,$keV band using $\alpha_X=1$.
To maintain consistency with the {\it ASCA} $\log N -
\log S$, the 0.3-3.5$\,$keV LF was extrapolated to
$L_X=10^{39}\,$\ergs, and $z<4$.  The predicted {\it ROSAT} QSO $\log
N - \log S$ based on the softer X-ray data lies a factor of two below
the {\it ASCA} QSO/narrow-emission-line-galaxy $\log N - \log S$ at
all fluxes, and suggests that the {\it ASCA} survey is detecting more
AGN than {\it ROSAT} at an equivalent flux limit, assuming a
straightforward extrapolation of the soft (0.5-2$\,$keV) X-ray
spectral slope $\alpha_X=1$ (Wilkes and Elvis 1987).  A harder
spectral slope would give greater agreement between the {\it ASCA} and
extrapolated {\it ROSAT} counts.

\section{Conclusions}\tx

We have obtained spectroscopic identification for a sample of 26 {\it
ASCA} sources detected in three deep {\it ROSAT} fields.  The optical
identification for the sample lies between 65 and 84 per cent.  Most
of the objects are identified as QSOs or narrow-emission-line
galaxies, with between 10 and 13 QSOs and between 1 and 5
narrow-emission-line galaxies in the final sample.  When we combine
this sample with a larger, predominantly low redshift ($z<0.2$)
sample, we find significant evidence for cosmological evolution, with
a luminosity evolution law $L_X\propto(1+z)^{2.04^{+0.16}_{-0.22}}$.
Based on the derived evolution and the 2-10$\,$keV luminosity
function, the predicted contribution of QSOs to the 2-10$\,$keV X-ray
background is $40\pm 20\,$per cent.  If the narrow-emission-line
galaxies identified in this analysis are also included (tentatively
identified as Seyfert 2 galaxies), then the total AGN
contribution to the X-ray background could amount to $80\pm 20$ per
cent. However, uncertainties over the extent to which the 
$z=0$ AGN LF can be extrapolated to low luminosities, could still
leave room for an as yet unidentified class of source to 
comprise a significant component of the 2-10$\,$keV XRB. 

If AGN are to prove the dominant source of the X-ray background, then
their average spectral properties must be consistent with that of the
background.  Currently, the results on the 2-10$\,$keV spectral slopes
of the {\it ASCA} AGN identified in this analysis are inconclusive and
improved X-ray observations are required for a more accurate estimate
of the 2-10$\,$keV spectral indecies of these objects.  Better optical
spectroscopy is also required to identify the true nature of the
narrow-emission-line galaxies.  

Finally, more faint and more high redshift $z>2$ AGN need to be
identified in the hard X-ray bands ($>2\,$keV).  This will give a
better estimate of the AGN LF in regions of luminosity-redshift space
where there is currently little information, and enable more accurate
estimates of the AGN contribution to the hard XRB to be made based on
the extrapolation of the AGN LF and its evolution with redshift.

\section*{Acknowledgements}\tx

\noindent  Optical spectroscopy was carried out with the 
Anglo-Australian Telescope, based on initial identifications of
candidates obtained from APM measurements of UK Scmhidt telescope
plates. We would like to thank Karl Glazebrook and Frank Freeman for
providing excellent assistance during the AAT run.  We would also like
to thank the referee, Xavier Barcons and Maite Ceballos for drawing
our attention to the LMA sample, and supplying appropriate data from
that sample.  We thank Bruce Grossan for providing a
copy of his thesis.  We are also indebted to Elihu Boldt for pointing
out a numerical error in an earlier version of this paper.
KFG acknowledges the receipt of a PPARC studentship.

\section*{References}
\bibitem
Almaini, O., 1996, Ph.D. thesis, University of Durham
\bibitem
Avni Y. Bahcall J.N. 1980, ApJ, 235, 694
\bibitem
Baldwin J.A., Phillips M.M., Terlevich R.J., 1981, PASP, 93, 5
\bibitem
Barcons X., Franceschini A., De Zotti G., Danese L., Miyati T.,
1995, ApJ, 455, 480
\bibitem
Boldt E., Leiter D., 1994, in Bicknell G.V., Dopita M.A., Quinn P.J.,
eds, ASP Conf. Ser. 54, The Physics of Active Galaxies, 
Astron. Soc. Pac., San Francisco, p131
\bibitem
Bower R.G. et al. 1996, MNRAS, 281, 59
\bibitem
Boyle B.J., 1996, Observatory, 116, 11
\bibitem
Boyle B.J., Shanks T., Peterson B.A., 1988, MNRAS, 235, 935
\bibitem
Boyle B.J., Staveley-Smith L., Stewart G.C., Georgantopoulos I., 
Shanks T., Griffiths R.E., 1993, MNRAS, 265, 501
\bibitem
Boyle B.J., Griffiths R.E., Shanks T., Stewart G.C., 
Georgantopoulos I., 1994, MNRAS, 271, 639
\bibitem
Boyle B.J., McMahon R.G., Wilkes B.J., Elvis M., 1995, MNRAS, 276, 315
\bibitem 
Carballo R., Warwick R.S., Barcons X., Gonzalez-Serrano J.I., Barber
C.R., Martinez-Gonzalez E., Perez-Fournon I., Burgos J., 1995, MNRAS,
277, 1312
\bibitem
Ceballos M.T., Barcons, X. 1996, MNRAS, 282, 493
\bibitem
Chen L.-W., Fabian A.C., Gendreau K.C., 1996, MNRAS, in press
\bibitem
Comastri A., Setti G., Zamorani G., Hasinger G., 1995, A\&A, 296, 1
\bibitem
Couch W.J., Shanks T., Pence, W.N., 1985, MNRAS, 213, 215
\bibitem
Fabian A.C., Barcons X., 1992, ARAA, 30, 429
\bibitem
Gendreau et al. 1995, PASJ, 47, L5
\bibitem
Georgantopoulos I., Stewart G.C., Shanks T., Boyle B.J., 
Griffiths R.E., 1996, MNRAS, 280, 276
\bibitem
Georgantopoulos I., Stewart G.C., Blair A.J., Shanks T., 
Griffiths R.E., Boyle B.J., Almaini O., Roche N., 1997, 
in press (Paper I)
\bibitem
Grossan B.A., 1992, PhD thesis, MIT
\bibitem
Hasinger G., Burg R., Giacconi R., Hartner G., Schmidt M., Trumper J., Zamorani G., 1993, AA, 275, 1
\bibitem
Maccacaro T., Della Ceca R., Gioia I.M., Morris S.L., Stocke J.T., 
Wolter A., 1991, ApJ, 374, 117
\bibitem
Madau P., Ghisellini, G., Fabian A.C., 1994, MNRAS, 270, L17
\bibitem
Metcalfe N., Shanks T., Fong R., Jones L.R., 1991, MNRAS, 249, 498
\bibitem 
Mushotsky R.F., 1982, Adv. Space Res., 3, 10
\bibitem 
Nandra K., Pounds K.A., 1994, MNRAS, 268, 405
\bibitem
Persic M., De Zotti G., Danese L., Palumbo G.G.C., Franceschini A.,
Boldt E.A., Marshall F.E. 1989, ApJ, 344, 125 
\bibitem
Peterson B.A., Ellis R.S., Efstathiou G., Shanks T., Bean A.J., 
Fong R., Zen-Long Z., 1986, MNRAS, 221, 233
\bibitem
Piccinotti G., Mushotzky R.F., Boldt E.A., Holt S.S., Marshall F.E.,
Serlemitsos P.J., Shafer R.A., 1982, ApJ, 253, 485
\bibitem
Pounds K.A., Nandra K, Stewart G.C., George I.M., Fabian A.C.,
1990, Nat, 344, 132
\bibitem
Remillard R.A., Bradt H.V., Buckley D.A.H., Roberts W.,
Schwarz D.A., Tuohy I.R., Wood K., 1986, ApJ, 301, 742
\bibitem
Schade D., Crampton D., Hammer F., Le Fevre O., Lilly S.J., 1996, 
MNRAS, 278, 95
\bibitem
Shanks T., Geogantopoulos I., Stewart G.C., Pounds K.A., Boyle B.J.,
Griffiths R.E., 1991, Nat, 353, 315
\bibitem
Tresse L., Rola C., Hammer F., Stasinska G., Le Fevre O., Lilly S.J.,
Crampton D., 1996, MNRAS, 281, 847
\bibitem 
Tucker W.H., Schwarz D.A., 1986, ApJ, 308, 53.
\bibitem
Wilkes B.J., Elvis M., 1987, ApJ, 323, 243
\bibitem 
Wood K.S. et al. 1984, ApJS, 56, 507
\bye